\documentclass[runningheads]{llncs}

\input{preamble}

\def\unit{I}

\newcommand\no[1]{\ensuremath{\overline{#1}}}

\newcommand{\init}{\mathit{init}}

\newcommand{\defn}{\,\triangleq\,}
\newcommand{\bool}{\ensuremath{\mathbb{B}}}
\newcommand\concept[1]{\textit{#1}}
\newcommand{\com}{\mathds{C}}
\newcommand\true{1}
\newcommand\false{0}

\newcommand\newl[1]{#1}
\newcommand\new[1]{#1}

\tikzstyle{leaf}=[draw, rectangle,minimum size=4.mm, inner sep=3pt]
\tikzstyle{var}=[circle,draw=black!70,solid,thick,minimum size=6mm]
\tikzstyle{bdd}=[regular polygon, regular polygon sides=3, draw=black!70,solid,thick,inner sep=0.5mm]
\tikzstyle{n}=[->,loosely dashed,thick]
\tikzstyle{p}=[->,solid,thick]
\tikzstyle{b}=[->,densely dashdotted,ultra thick]
\tikzstyle{e0}[0]=[dashed,thick,bend right=#1]
\tikzstyle{e1}[0]=[solid, bend left =#1]

\tikzstyle{lbl}=[draw,fill=white,inner sep=2pt, minimum size=0cm,line width=.5pt]

\makeatletter
\newcommand*\circled[2][1.6]{\tikz[baseline=(char.base)]{
    \node[shape=circle, draw, inner sep=1pt, 
        minimum height={\f@size*#1},] (char) {\vphantom{WAH1g}#2};}}
\makeatother

\makeatletter
\newcommand*\circledd[3][1.6]{\tikz[baseline=(char.base)]{
    \node[shape=circle, draw, inner sep=1pt, 
        minimum height={\f@size*#1},label=below:{\scriptsize #3}] (char) {\vphantom{WAH1g}#2};}}
\makeatother

\def\bigarrowhead{-{Latex[length=2mm,width=2mm]}}

\undef\exercise
\newcommand\exercise[1]{
\tcbset{enhanced,colback=green!5!white,colframe=green!75!black,fonttitle=\bfseries}
\begin{tcolorbox}
\textbf{Exercise}: #1
\end{tcolorbox}
}

\undef\explainerbox
\newcommand\explainerbox[2]{
\tcbset{enhanced,colback=yellow!5!white,colframe=yellow!75!black,fonttitle=\bfseries}
\begin{tcolorbox}[
    breakable,
    skin first=enhanced,
    skin middle=enhanced,
    skin last=enhanced,
    ]
\textbf{#1} #2
\end{tcolorbox}
}

\def\stwo{\frac{1}{\sqrt{2}}}
\def\tstwo{\tfrac{1}{\sqrt{2}}}

\newcommand\keymessage[2][]{
\tcbset{enhanced,colback=cyan!5!white,colframe=cyan!75!black,fonttitle=\bfseries}
\begin{tcolorbox}
\textbf{Key message#1}: #2
\end{tcolorbox}
}
\makeatletter
\newcommand{\printfnsymbol}[1]{
  \textsuperscript{\@fnsymbol{#1}}
}
\makeatother

\newcommand{\aref}[1]{\hyperref[#1]{Appendix~\ref*{#1}}}

\begin{document}

\title{Advancing Quantum Computing with Formal Methods}
\author{
Arend-Jan Quist\thanks{These authors contributed equally.}\orcidID{0000-0002-6501-2112}
, 
Jingyi Mei\printfnsymbol{1}\orcidID{0000-0002-4665-9818}, 
\\
Tim Coopmans\printfnsymbol{1}\orcidID{0000-0002-9780-0949}(\Letter),
Alfons Laarman\orcidID{0000-0002-2433-4174}
}
\authorrunning{Arend-Jan Quist, Jingyi Mei, Tim Coopmans, Alfons Laarman}
\date{April 2024}
\institute{
Leiden University, 
Leiden, The Netherlands\\ 
\email{\{a.quist,j.mei,t.j.coopmans,a.w.laarman\}@liacs.leidenuniv.nl}}

\maketitle

\begin{abstract}
This tutorial introduces quantum computing with a focus on the applicability of formal methods in this relatively new domain. We describe quantum circuits and convey an understanding of their inherent combinatorial nature and the exponential blow-up that makes them hard to analyze. Then, we show how weighted model counting (\#SAT) can be used to solve hard analysis tasks for quantum circuits.

This tutorial is aimed at everyone in the formal methods community with an interest in quantum computing. Familiarity with quantum computing is not required, but basic linear algebra knowledge (particularly matrix multiplication and basis vectors) is a prerequisite. The goal of the tutorial is to inspire the community to advance the development of quantum computing with formal methods.

\end{abstract}

\section{Introduction}
\label{sec:intro}
\vspace{-1em}

\begin{quote}
\textit{``Nature isn't classical, (...), and if you want to make a simulation of nature, you'd better make it quantum mechanical, and by golly it's a wonderful problem, because it doesn't look so easy''}\end{quote}
\hfill 
Richard~Feynman~\cite{feynman1982simulating}

Renowned physicist and Nobel prize winner Richard Feynman is said to be
the first to coin the idea of a `quantum computer'
by inverting the difficulty he encountered when solving equations in quantum mechanics~\cite{feynman1982simulating}.
He proposed to make nature's complexity work for us, instead of trying to calculate nature's behavior using classical tools ---which Feynman noticed requires performing exponentially many calculations.
Fast forward to today, and the first quantum computers have been built 
arguably showing first signs of `quantum advantage'~\cite{arute2019quantum}.

We see that powerful techniques from formal methods are ideally suited to tackle some of the crucial problems on the road towards a full-scale quantum advantage.
This tutorial, therefore, provides a crash course into quantum computing, specifically geared towards a formal methods audience. Our goal is to inform the community of the challenges in handling quantum computations and point them in the right direction to apply their own favorite techniques on open problems in the field of quantum computing (\autoref{sec:outro}).

As examples and for inspiration, we explain two approaches for analyzing quantum circuits: 
\#SAT and decision diagrams (DD).  Both are formalisms with which the formal methods community is well familiar.
We show how to formulate the semantics of quantum circuits using both.
At the same time, we keep the quantum background to an absolute minimum by explaining quantum computing as a minor extension of reversible computation and probabilistic computation,\footnote{Surprisingly, not even complex numbers are needed, as the (classical) reversible Toffoli gate, which is universal for reversible computing, requires only an additional Hadamard (Walsh-Hadamard / Reed-Muller transform) gate to yield a universal gate set for quantum computation~\cite{aharonov2003simple,shi}. We discuss other gate sets in \aref{app:gatesets}.}
 and gradually introducing the required notation.

Nevertheless, our basic exposition manages to touch upon quantum algorithms, such as Grover search~\cite{grover1996fast}, by briefly discussing how the Boolean satisfiability problem can be incorporated into a quantum circuit as an oracle. While the satisfying assignments can then be found in the so-called \concept{probability amplitudes} of the quantum state that is computed by the circuit, it is hard to extract such information using \concept{measurements}, the only method to extract physical information from a quantum system (which is unfortunately rather crude).
Such an example should inspire listeners to think about how to extract the solution to the satisfiability problem using measurements, which is the difficulty of coming up with so-desired quantum algorithms that outperform their classical alternatives.\footnote{While there is, for instance, an efficient quantum algorithm for factoring (Shor's~\cite{shor1994algorithms}), we do not know for sure whether factoring is hard classically. So a true \concept{separation} between the classical and quantum complexity classes is still open (formalized as the $\BPP = \BQP$ question).}

Our exposition here focuses on the task of (classically) simulating quantum circuits, which is formally defined in the background section. This ensures that the audience becomes familiar with the semantics of quantum circuits and algorithms (which can be formulated as uniform families of quantum circuits). At the very end, we show, based on related works, how the discussed encodings in \#SAT can also be used to solve various other tasks, inter alia: (quantum circuit) equivalence checking, synthesis, optimization and quantum Hoare logic checking. We also point the audience in the direction of the current obstacles on the road to quantum supremacy, such as, circuit optimization and quantum error correction (which will crucially realize the ideal quantum circuit model introduced here). Here, we even point out connections beyond quantum computing since it is easy to show that progress in handling quantum circuits translates to progress in solving important and hard problems in quantum mechanics, like simulation of many-body physics and finding ground states.

\section{Quantum computing for computer scientists}
\label{sec:quantumComputingFromCSperspective}
\label{sec:prelims}
\label{sec:preliminaries}

\begin{quote}
\textit{``Quantum computing becomes easy once you take the physics out of it''}
\end{quote}
\hfill \url{https://scottaaronson.blog}

This section introduces quantum circuits as a generalization of (classical) reversible and probabilistic computing.
Having provided the reader with an understanding of these classical building blocks, we then explore the semantics of quantum circuits.
We finish by highlighting the crucial aspects of a quantum circuit that make formal methods amenable to it.

\subsection{From reversible, via probabilistic, to quantum circuits}
\label{sec:background-intro}
Each (irreversible) Boolean circuit can be written as a reversible circuit.
One way to do this is using circuits with only the so-called Toffoli gate~\cite{toffoli1980reversible} (see the yellow box below).

\explainerbox{Toffoli gate.}{
{
The three-bit Toffoli gate, shown below, is both reversible and classically universal. 
It takes three bits $a,b,c$ as inputs and outputs $a, b, c \oplus ab$. Informally, it only flips the $c$ bit when both $a$ and $b$ are true (leaving $a$ and $b$ intact).
\vspace{-1em}
\[
\scalebox{1}{
\Qcircuit @C=1em @R=.2em @!R {
\lstick{a} & \ctrl{2}  &\qw 	& \rstick{\hspace{-1em}a' := a} \\
\lstick{b} & \ctrl{1}  &\qw	    & \rstick{\hspace{-1em}b' := b} \\
\lstick{c} & \targ     &\qw     & \rstick{\hspace{-1em}c' := c \oplus (a\land b)}  \\  
}}
\]
Applying a Toffoli twice will compute $c \oplus ab \oplus (a\land b) = c$, thus reversing the original computation by `uncomputing' the result.
Furthermore, by setting $a=b=1$, the $c$ bit is always flipped so that we can realize a NOT gate ($\neg$) as syntactic sugar (drawn as \raisebox{.0em}{$ \Qcircuit @C=.5em @R=.2em @!R {
& \targ  &\qw \\
}$} in a circuit). 
Toffoli is classically universal, 
which follows from the fact that the gate set $\set{\land, \neg}$ is universal and that Toffoli implements an AND gate ($\land$).
By setting $a=1$, we obtain a (singly-)controlled-NOT, or CNOT gate (drawn as
\raisebox{.7em}{$ \Qcircuit @C=.4em @R=.2em @!R {
& \ctrl{1}  &\qw \\
& \targ     &\qw \\  
}$
}), which 
flips the value of $c$ only when $b$ is set.
As additional syntactic sugar, we use the negated control $\Qcircuit @C=.4em @R=.2em @!R { & \ctrlo{0} & \qw }$, which is a control $\Qcircuit @C=.4em @R=.2em @!R { & \ctrl{0} & \qw }$ surrounded by negations $\Qcircuit @C=.5em @R=.2em @!R { &\targ & \ctrl{0} &\targ &\qw }$. It checks whether the input is false, returning it to its original state.
}
}

As an example of writing each Boolean circuit as a reversible circuit, consider circuit (i) on input $a, b, c\in \{0, 1\}$ below: We construct a reversible circuit by storing intermediate value $x$ and the result $y$ separately, adding wires for these variables. This leads to the reversible circuit (ii) containing NOT gates (drawn as $\oplus$ indicating a bit flip without control) and Toffoli gates (a bit flip $\oplus$ with two controls \raisebox{.3mm}{
$
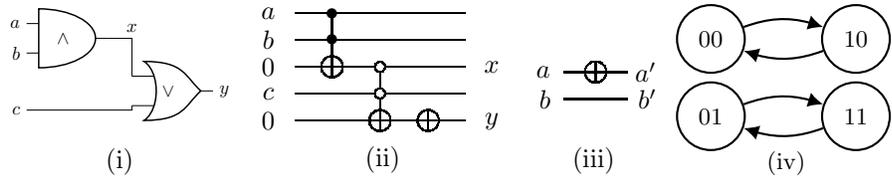
\Qcircuit @C=.4em @R=.2em @!R { & \ctrl{0} &\qw}$} or  \raisebox{.3mm}{$\Qcircuit @C=.4em @R=.2em @!R { & \ctrlo{0} &\qw}$}).

\mbox{ } 
\begin{minipage}{.28\textwidth}
\scalebox{0.7}{
\begin{circuitikz}[node distance=.2cm]
\draw
(0,2) node[and port] (myand) {\hspace{-1ex}$\land$}
(2,1) node[or port] (myor) {\hspace{-1ex}$\lor$}
(myand.in 1) node[anchor=east] {$a$}
(myand.in 2) node[anchor=east] (bnode) {$b$}
(myand.out) -| node[above]{$x$} (myor.in 1)
node[below=.638cm of bnode] (cnode) {$c$}
(cnode) -|  (myor.in 2)
(myor.out) node[anchor=west] {$y$}
;
\end{circuitikz}
}
\centering\\
\vspace{.2cm}
(i)
\end{minipage}
\begin{minipage}{.28\textwidth}
\scalebox{1}{
$
\Qcircuit @C=1em @R=.2em @!R {
{a} & &  \ctrl{2} 	& \qw      & \qw  & \qw & \\
{b} & &  \ctrl{1}	& \qw      & \qw  & \qw & \\
{0} & & \targ       & \ctrlo{1} & \qw & \qw & {x}\\  
{c} & &     \qw     & \ctrlo{1} & \qw  & \qw & \\
{0} &  &     \qw    & \targ & \targ  & \qw & {y} \\
}$
}
\newline
\mbox{ }
\newline
\centering
(ii)
\end{minipage}~~~~

\hspace{-.2cm}
\begin{minipage}{.25\textwidth}
\mbox{ }
\newline
\mbox{ }
\newline
\mbox{ }
$
\scalebox{1}{
\Qcircuit @C=.8em @R=.2em @!R {
\lstick{a} &   \targ{} 	& \qw  & \push{} & \lstick{~a'}\\
\lstick{b} &   \qw		& \qw  & \push{} & \lstick{~b'}\\
}}
$
\newline
\mbox{ }
\newline
\mbox{ }
\newline
\mbox{ } ~~(iii)
\end{minipage}
\begin{minipage}{.25\textwidth}
\hspace{-1.4cm}
\begin{tikzpicture}[
    scale=0.3,
    every path/.style={>=latex},
    every node/.style={},
    inner sep=0pt,
    minimum size=14pt,
    line width=1pt,
    node distance=.5cm,
    thick,
    font=\footnotesize
    ]

    \node[draw, circle, minimum size=0.9cm] (00)   {$00$};
    \node[draw,circle, right =1cm of 00, xshift=0cm, minimum size=0.9cm] (10) {$10$};
    \node[draw,circle, below =.1cm of 00, xshift=0cm, minimum size=0.9cm] (01) {$01$};
    \node[below = -.3cm of 01, xshift=1cm, minimum size=0.9cm] {(iv)};
    \node[draw,circle, below =.1cm of 10, xshift=0cm, minimum size=0.9cm] (11) {$11$};

\draw[black, \bigarrowhead, bend left=20] (00) edge node[yshift=0.2cm] {} (10);
\draw[black, \bigarrowhead, bend left=20] (10) edge node[yshift=-0.2cm] {} (00);
\draw[black, \bigarrowhead, bend left=20] (01) edge node[yshift=0.2cm] {} (11);
\draw[black, \bigarrowhead, bend left=20] (11) edge node[yshift=-0.2cm] {} (01);

\end{tikzpicture}
\end{minipage}

Next, let us visualize the action of a classical reversible circuit.
In (iii) above, we use a circuit on two bits $a, b$ and give in (iv) an automaton of the four possible states on two bits.
An arrow from state $(a, b)$ to $(a', b')$ indicates that the circuit maps $(a, b)$ to $(a', b')$.
(We use a circuit on two bits instead of five as in (ii) to avoid too large automata for easy readability.)

\exercise{Create a reversible circuit for the Boolean formula $((a\land b) \lor c) \land d$. Then extend the circuit to uncompute all wires but the result.}

\vspace{-1\baselineskip}
\subsubsection{Probabilistic classical (reversible) computing with linear algebra.
\label{sec:probabilistic-computing}
}

To move closer towards quantum circuits, we will now see how the automaton changes when the logical gates used are probabilistic.
Specifically, in example (iii) above, we replace the NOT gate ($\oplus$) with a probabilistic gate $G$, which does nothing with probability $75\%$ and applies a NOT with $25\%$, resulting in (v) below.
Then the circuit is modeled by a Markov chain, and the transition arrows in the automaton (vi) are labeled with the probabilities of mapping the input state $(a, b)$ to output state $(a', b')$~\cite{kemeny1969finite}.

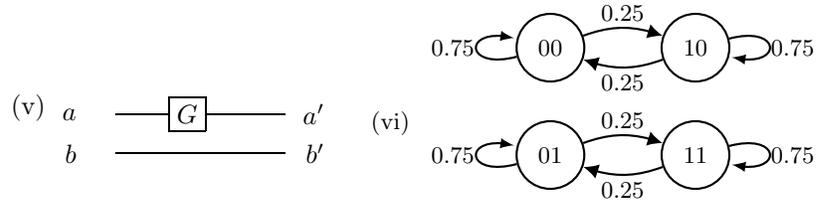
\begin{minipage}{.45\textwidth}

\begin{tikzpicture}[
    scale=0.3,
    every path/.style={>=latex},
    every node/.style={},
    inner sep=0pt,
    minimum size=14pt,
    line width=1pt,
    node distance=.5cm,
    thick,
    font=\footnotesize
    ]
    \node[yshift=0cm, minimum size=0.9cm] {};
\end{tikzpicture}
(v)
\quad
$
\scalebox{1}{
\Qcircuit @C=1em @R=.2em @!R {
\lstick{a} & & \qw &  \gate{G} 	&\qw	& \qw      & \qw  & \push{} & \lstick{a'}\\
\lstick{b} & & \qw &  \qw	&\qw 	& \qw      & \qw  & \push{} & \lstick{b'}\\
}}
$
\newline
\mbox{ }
\newline

\end{minipage}
\begin{minipage}{.45\textwidth}
\begin{tikzpicture}[
    scale=0.3,
    every path/.style={>=latex},
    every node/.style={},
    inner sep=0pt,
    minimum size=14pt,
    line width=1pt,
    node distance=.5cm,
    thick,
    font=\footnotesize
    ]

    \node[draw, circle, minimum size=0.9cm] (00)   {$00$};
    \node[left=1.2cm of 00, yshift=-1cm, minimum size=0.9cm] {(vi)};
    \node[draw,circle, right =1cm of 00, xshift=0cm, minimum size=0.9cm] (10) {$10$};
    \node[draw,circle, below =.5cm of 00, xshift=0cm, minimum size=0.9cm] (01) {$01$};
    \node[draw,circle, below =.5cm of 10, xshift=0cm, minimum size=0.9cm] (11) {$11$};

\draw[black, \bigarrowhead, bend left=20] (00) edge node[yshift=0.2cm] {$0.25$} (10);
\draw[loop left, black, \bigarrowhead] (00) edge node[yshift=0.0cm] {$0.75$} (00);

\draw[black, \bigarrowhead, bend left=20] (10) edge node[yshift=-0.2cm] {$0.25$} (00);
\draw[loop right, black, \bigarrowhead] (10) edge node[yshift=0.0cm] {$0.75$} (10);

\draw[black, \bigarrowhead, bend left=20] (01) edge node[yshift=0.2cm] {$0.25$} (11);
\draw[loop left, black, \bigarrowhead] (01) edge node[yshift=0.0cm] {$0.75$} (01);

\draw[black, \bigarrowhead, bend left=20] (11) edge node[yshift=-0.2cm] {$0.25$} (01);
\draw[loop right, black, \bigarrowhead] (11) edge node[yshift=0.0cm] {$0.75$} (11);
\end{tikzpicture}
\newline
\end{minipage}

A Markov chain is typically analyzed by associating each state $(a,b)$ (abbreviated `$ab$') in the automaton to a basis vector and writing down the transition probabilities as an adjacency matrix~$M$ of the automaton above:
\begin{equation}
\label{eq:compbasvectors}
`00\text' \equiv\hspace{-.5mm} \begin{bmatrix} 1 \\0\\0\\0\end{bmatrix}\hspace{-1mm},
`01\text' \equiv \begin{bmatrix} 0 \\1\\0\\0\end{bmatrix}\hspace{-1mm},
`10\text' \equiv \begin{bmatrix} 0 \\0\\1\\0\end{bmatrix}\hspace{-1mm},
`11\text' \equiv \begin{bmatrix} 0 \\0\\0\\1\end{bmatrix}\hspace{-1mm},~
M = 
\begin{bmatrix}
0.75& 0 & 0.25 & 0\\
0 & 0.75 & 0 & 0.25\\
0.25 & 0 & 0.75 & 0\\
0 & 0.25 & 0 & 0.75
\end{bmatrix}
\end{equation}

Next, the probability distribution over the output states $(a', b')$ on input $(a, b)$ is computed using matrix-vector multiplication.
For example, on input `00' $(a=0, b=0)$, the output is
\[
M\cdot `00\text' =
\begin{bmatrix}
0.75& 0 & 0.25 & 0\\
0 & 0.75 & 0 & 0.25\\
0.25 & 0 & 0.75 & 0\\
0 & 0.25 & 0 & 0.75
\end{bmatrix} \cdot 
\begin{bmatrix}1\\0\\0\\0\end{bmatrix} = \begin{bmatrix}0.75\\0\\0.25\\0\end{bmatrix}
=
0.75 \begin{bmatrix}1\\0\\0\\0\end{bmatrix}
+
0.25 \begin{bmatrix}0\\0\\1\\0\end{bmatrix}
,
\]
that is, `$00$' is mapped to itself with probability $75\%$ and to `$10$' with probability $25\%$, as was already revealed by the Markov chain.

The advantage of using the transition matrix instead of writing down the Markov chain is that the input need not be a single bitstring itself but can also be a probability distribution over bitstrings. In this more general case, \textbf{matrix-vector multiplication correctly propagates the probabilities.}
To see this, we first directly compute the probabilities of the output bitstrings for the circuit in (v) using the Markov chain.
We use the example of an input state that is either set to the bitstring `$00$' with probability $0.4$ or to the bitstring `$10$' with probability $0.6$.
In the first case, the output of the circuit in (v) is `$00$' with a probability of $0.75$ and `$10$' with a probability of $0.25$; in the second case, it is `$00$' (`$10$') with a probability of $0.25$ ($0.75$).
Aggregating these two, the output is `$00$' with probability $0.4 \cdot 0.75 + 0.6 \cdot 0.25 =0.45$ and `$10$' with probability $0.4 \cdot 0.25 + 0.6 \cdot 0.75 = 0.55$.
Equivalently, we obtain this result using linear algebra.
The input, which represents `$00$' with probability $0.4$ and `$10$' with probability $0.6$, is the vector $\vec{v} = \begin{bmatrix} 0.4& 0 &0.6&0\end{bmatrix}^{\top}$ (where the transpose $(.)^{\top}$ turns a row vector into a column vector).
The output probability distribution is computed by applying the transition matrix to the input vector $\vec{v}$:
\[
M \cdot \vec{v} = \begin{bmatrix} 0.4 \cdot 0.75 + 0.6 \cdot 0.25\\ 0\\ 0.4 \cdot 0.25 + 0.6 \cdot 0.75\\0\end{bmatrix} = \begin{bmatrix} 0.45\\ 0 \\ 0.55\\0 \end{bmatrix},
\]
representing 45\% in state `$00$' and $55\%$ in state `$10$', which is precisely the same result we obtained before!
Indeed, matrix-vector multiplication correctly propagates the probabilities.

We note that because the input vector represents a probability distribution over bitstrings, its entries should be probabilities, i.e., nonnegative values between zero and 1, and the vector should be \concept{normalized} (all its entries should sum to 1). Also, each column in the transition matrix has the same property, as the sum of outgoing-edge labels of a node in the automaton constitute a probability distribution themselves too.

\keymessage{An $n$-bit reversible circuit with probabilistic gates is represented by a transition matrix. If the input to the circuit is a probability distribution over bitstrings, it can be represented by a $2^n$-length vector containing the probabilities. Furthermore, the action of the circuit is computed by applying the transition matrix to the input vector.
}

Finally, in order to compute the probability of ending in a certain state, say `10', after starting in $\vec v$, we can compute the following:
`$10\text'^\top \cdot M \cdot \vec v  \equiv \begin{bmatrix} 0 & 0 & 1 & 0\end{bmatrix} \cdot M \cdot \vec v = 0.55$.
Note the use of a row vector (`$10\text'^\top$) and a column vector ($\vec v$).

\emph{Sequential composition of circuits using matrix multiplication.}
Suppose that the input probability distribution over bitstrings $\vec{v}$ is inputted to circuit $A$, yielding the output vector $\vec{w} = M_A \vec{v}$, where $M_A$ is the transition matrix of $A$.
Suppose furthermore that we pass input $\vec{w}$ to another circuit $B$.
Then the probability vector representing $B$'s output is $M_B \vec{w} = M_B(M_A \vec{v})$, which can also be written as $(M_B\cdot M_A)\vec{v}$, where $\cdot$ denotes matrix multiplication.
Since this holds for any input vector $\vec v$, we observe that a circuit $C$, consisting of first performing $A$ followed by $B$, has a transition matrix $M_B \cdot M_A$.
That is, the sequential composition of gates, and thus of circuits, corresponds to the matrix multiplication of their transition matrices.
For example, in (v) the transition matrix is denoted as $M$ and $G$ is applied to the top bit $a$ once, so applying $G$ twice instead yields~$M^2$:
\[
~~\scalebox{1}{
\raisebox{.3cm}{
\Qcircuit @C=.6em @R=.2em @!R {
\lstick{a} & \qw &  \gate{G} 	&\gate{G}	& \qw      & \qw  & \rstick{a'}\\
\lstick{b} & \qw &  \qw	&\qw 	& \qw      & \qw  &  \rstick{b'}\\
}}}
\quad~~~~
M^2 =
\begin{bmatrix}
0.75& 0 & 0.25 & 0\\
0 & 0.75 & 0 & 0.25\\
0.25 & 0 & 0.75 & 0\\
0 & 0.25 & 0 & 0.75
\end{bmatrix}^2
=
\begin{bmatrix}
0.625 & 0 & 0.375 & 0\\
0 & 0.625 & 0 & 0.625\\
0.375 & 0 & 0.375 & 0\\
0 & 0.375 & 0 & 0.625
\end{bmatrix}
\]

\emph{Parallel composition of circuits using the Kronecker product.}
Above, we found that the sequential composition of circuits corresponds to the matrix-multiplication product of their transition matrices.
For parallel composition, where one stacks two circuits `on top' of each other, we need the Kronecker product of matrices.
\explainerbox{Kronecker product.}{
The Kronecker product of matrices $A$ and $B$ performs a giant case distinction: for each entry $a$ of $A$, we create a copy of all possible entries $b$ of $B$, and the resulting matrix contains the values $a \cdot b$.
For example, the transition matrices on a single bit
\[
A = \begin{bmatrix}
0.1 &~~ 0.2\\
0.9 &~~ 0.8
\end{bmatrix}
,~~~~
B = \begin{bmatrix}
0.3 &~~ 0.4\\
0.7 &~~ 0.6
\end{bmatrix}
\]
is
\[
A\otimes B = 
\begin{bmatrix}
0.1 \cdot B &~~ 0.2 \cdot B\\
0.9 \cdot B &~~ 0.8 \cdot B
\end{bmatrix}
=
\begin{bmatrix}
0.1 \cdot 0.3 &~~ 0.1 \cdot 0.4 &~~ 0.2 \cdot 0.3 &~~ 0.2 \cdot 0.4\\
0.1 \cdot 0.7 &~~ 0.1 \cdot 0.6 &~~ 0.2 \cdot 0.7 &~~ 0.2 \cdot 0.6\\
0.9 \cdot 0.3 &~~ 0.9 \cdot 0.4 &~~ 0.8 \cdot 0.3 &~~ 0.8 \cdot 0.4\\
0.9 \cdot 0.7 &~~ \underline{0.9 \cdot 0.6} &~~ 0.8 \cdot 0.7 &~~ 0.8 \cdot 0.6
\end{bmatrix}.
\]
For example, the entry of $A \otimes B$ at column 2 ($01 = 0_A 1_B$ in binary) and row 4 ($1_A 1_B$), which is underlined above, contains the probability $0.9 \cdot 0.6$ which is the probability that \emph{both} $A$ maps bit $0$ to $1$ (which happens with probability $0.9$) as well as $B$ maps $1$ to $1$ (which happens with probability $0.6$).\\

Formally, the Kronecker product on two general matrices (not necessarily transition matrices) acts as follows: given $r_A \times c_A$ matrix $A$ and $r_B \times c_B$ matrix $B$,
the $r_A r_B \times c_A c_B$ matrix $A \otimes B$ is
\[
A \otimes B = 
\begin{bmatrix}
A_{00} B & A_{01}B &\dots& A_{0 c_A} B\\
A_{10} B & A_{11}B &\dots& A_{1 c_A} B\\
\vdots&\vdots&\ddots&\\
A_{r_A 0} B & A_{r_A 1}B &\dots& A_{r_A c_A} B\\
\end{bmatrix}
.
\]
}

The product behavior of the Kronecker product of two matrices $A$ and $B$ can be interpreted as choosing an entry $a$ from $A$ and an entry $b$ from $B$ and multiplying them (see the yellow box above).
When $A$ and $B$ are probability vectors of the individual systems, their Kronecker product yields precisely the vector on the combined system, where the values $a\cdot b$ are the occurence probabilities of these automaton states/unit vectors.
The same interpretation shows that the parallel composition of two gates is also given by their Kronecker product.
For example, the transition matrix for $G$ in (iii) (which acts on a single bit) is given by $M_G = \begin{bsmallmatrix} 0.75 & 0.25 \\ 0.25 & 0.75\end{bsmallmatrix}$, the transition matrix on the second bit (doing nothing) is the identity matrix $\unit =  \begin{bsmallmatrix} 1 & 0 \\ 0 & 1\end{bsmallmatrix}$; hence we may write $M_G \otimes \unit$ for the transition matrix on the two bits $a, b$.

\exercise{Using the definition of the Kronecker product sign $\otimes$ (see the yellow box above), verify that $M_G \otimes \unit = M$ from \autoref{eq:compbasvectors}.}

\exercise{
For a single bit, the two basis vectors are `$0$' $\equiv \left[\begin{smallmatrix}1\\0\end{smallmatrix}\right]$ and `$1$'$\equiv \left[\begin{smallmatrix}0\\1\end{smallmatrix}\right]$.
Compute the $4$ possible Kronecker products between these states (`$0$' $\otimes$ `$0$', `$0$ $\otimes$ `$1$', etc.) and verify that these are precisely the four basis vectors for two bits from \autoref{eq:compbasvectors}.
}

\keymessage{
Composing probabilistic circuits sequentially corresponds to multiplying their transition matrices, while their parallel composition is represented by the Kronecker product of the transition matrices.}

\subsubsection{Quantum computations with linear algebra.}\label{sub:quantum}
Now let us move to quantum bits or qubits. \emph{Here, we limit the presentation to a simplified model of quantum computing using solely real numbers. This model is universal for quantum computing~\cite{aharonov2003simple} and suffices to explain all aspects of quantum computing. Nevertheless, a reader should keep in mind that the usual quantum computing model contains complex numbers and, therefore, comprises the subtle differences noted below (also see reading material referenced at the end of \autoref{sec:outro}).}

The state of two qubits is a generalization of a probability distribution over the four bitstrings, whose vectors, e.g., for `00', `01', `10' and `11'
in \autoref{eq:compbasvectors}, are referred to as the \emph{computational-basis states} as they form a basis of the space of $4$-dimensional vectors, and written as $\ket{00}, \ket{01}, \ket{10}, \ket{11}$ (so-called Dirac notation).
Just like in the Markov chain case above, the state on $n$ quantum bits is represented by a vector of length $2^n$.
However, in contrast to the Markov chain case, the entries of the vector are not probabilities (which sum up to $1$) but are so-called \emph{probability amplitudes}, or simply \concept{amplitudes}. 
So to represent a quantum state, the vector should now be normalized by letting the \emph{squares}\footnote{Or modulus square in the general case with complex numbers.} of these amplitudes sum up to $1$.
Instead of a transition matrix, which contains probabilities such that the entries in each column add up to 1,
the gate in the circuit is given by a \emph{unitary matrix}: in a unitary matrix, the sum of \emph{squares} of the entries (amplitudes) in a column add up to 1.\footnote{The rows of a unitary matrix are also orthogonal; see \autoref{sec:quantum-circuit-building-blocks} for full definition.}
Sequential and parallel composition of gates, identically to the case of probabilistic computing in \autoref{sec:probabilistic-computing}, correspond to matrix multiplication and Kronecker product, respectively.

An example is the circuit below: the input is computational-basis state \newl{$\ket a \otimes \ket b = \ket{ab}$ for $a, b \in \{0, 1\}$,} and the single-qubit quantum gate $U$ in the circuit yields the two-qubit action $M'$ (to see why $M' = U\otimes \unit$, compare with $M_G \otimes I$ in \autoref{sec:probabilistic-computing}):
\[
~~~~\raisebox{1em}{
\scalebox{1}{
\Qcircuit @C=.5em @R=.2em @!R {
\lstick{\ket a}
& \qw &  \gate{U} 	&\qw & \\
\lstick{\ket b}
& \qw &  \qw	&\qw 
\\
}
}}
~~
U = 
\begin{bmatrix}
\sqrt{0.75}& \sqrt{0.25}\\
\sqrt{0.25} & -\sqrt{0.75}\\
\end{bmatrix}
\quad
M' = 
U \otimes \unit
=
\left[
\begin{smallmatrix}
\sqrt{0.75}& 0 & \sqrt{0.25} & 0\\
0 & \sqrt{0.75} & 0 & \sqrt{0.25}\\
\sqrt{0.25} & 0 & -\sqrt{0.75} & 0\\
0 & \sqrt{0.25} & 0 & -\sqrt{0.75}
\end{smallmatrix}
\right]
\]
\\
For example, if the input state is $\ket{11} \equiv \begin{bmatrix} 0, 0, 0, 1\end{bmatrix}^{\top}$, then the output state is $(U\otimes \unit) \cdot \ket{11} = \begin{bmatrix} 0, \sqrt{0.25}, 0, -\sqrt{0.75}\end{bmatrix}^{\top} = \sqrt{0.25} \ket{01} - \sqrt{0.75} \ket{11}$.
By squaring the amplitudes $\sqrt{0.25}$ and $\sqrt{0.75}$, we regain the interpretation as probability distribution over the states $\ket{01}$ and $\ket{11}$.
\emph{However}, the fact that the amplitudes can be \emph{negative} (and complex numbers in general) is crucial: as we will see later, these give rise to \emph{interference}, where amplitudes are canceled or amplified, a feature that is not present in probabilistic computing.

\keymessage{quantum computing is a generalization of probabilistic computing.

It will be useful to read a(n exponential-length) state vector as a probability (amplitude) distribution over classical (computational basis) states:
\[
\begin{bmatrix}\alpha_{0}\\\alpha_{1}\\\vdots\\\vdots\\ \alpha_{2^n-2}\\ \alpha_{2^n-1}\end{bmatrix}
~~=~~
\begin{bmatrix}\alpha_{0\dots00}\\\alpha_{0\dots 01}\\\vdots\\\vdots\\ \alpha_{1\dots10}\\ \alpha_{1\dots11}\end{bmatrix}
\begin{matrix} 
    ~~\rightarrow~~ \text{probability amplitude of } \ket{0\dots00}\\
    ~~\rightarrow~~ \text{probability amplitude of } \ket{0\dots01}\\
    \\
    \\
    \\
    ~~\rightarrow~~ \text{probability amplitude of }\ket{1\dots10}\\
    ~~\rightarrow~~ \text{probability amplitude of }\ket{1\dots11}\\
    \end{matrix}
\]
}

\subsection{Building blocks of quantum circuits \label{sec:quantum-circuit-building-blocks}}

In the previous section, we gave an example of a quantum state as a generalization of a probability distribution over bitstrings.
Here we give the full definition of a general quantum state (state of a collection of qubits), and give the building blocks of quantum-computer circuits: the quantum analogs of logical \emph{gates}, and \emph{measurements} as a means to extract information from a quantum state. (For a full introduction, see \cite{nielsen2000quantum}.)

\subsubsection{Quantum bits}
As we have seen before, a quantum state on $n$ qubits is a column vector of length $2^n$, which constitutes a probability distribution over the length-$n$ bitstrings $b \in \{0, 1\}^n$ as the vector is a \emph{linear combination} $\sum_{b \in \{0, 1\}^n} \alpha_{b} \ket{b}$ over the corresponding computational-basis states $\ket{b}$ (i.e. the length-$2^n$ vector with entry $1$ at position $b$ and $0$ everywhere else).
(The quantum-computing jargon is \emph{superposition} instead of linear combination.)
The vector entries (amplitudes) $\alpha_b$ are complex numbers in general. In this paper, we will always consider real vector entries.
The square of the amplitude determines the probability.
For example, the following vectors
\begin{equation}
\label{eq:states}
\begin{bmatrix} -\sqrt{0.75}\\\sqrt{0.25}\end{bmatrix}
,\quad
\begin{bmatrix} -\sqrt{0.75}\\-\sqrt{0.25}\end{bmatrix}
,\quad
\begin{bmatrix} \sqrt{0.75}\\\sqrt{0.25}\end{bmatrix}
\end{equation}
all give rise to the probabilities $0.75$ and $0.25$ over the states $\ket{0}$ and $\ket{1}$, respectively.
If the amplitudes $\alpha_b$ of a vector $\vec{v}$ are real numbers, then $\sqrt{\sum_{b\in \{0, 1\}^n} \alpha_b^2}$ indicates the vector's length
(for complex numbers, we would also have to take the modulus $|\alpha_b|$ instead of $\alpha_b$).
Since the values $|\alpha_b|^2$ form a probability distribution, we find that:

\keymessage{an $n$-qubit state is a vector of $2^n$ complex numbers with norm 1. (In this paper, we only consider real vectors.)}

\begin{example}
\label{ex:states}
The vectors $\ket{\phi}, \ket{\psi}$ and $\ket{\eta}$ are quantum states on $2$, $2$ and $3$ qubits, respectively:
\[
\ket{\phi} \defn
\tfrac{1}{\sqrt{1^2 + 2^2 + 3^2 + \sqrt{17}^2}} \cdot \begin{bmatrix} 1\\ 2\\ -3\\ \sqrt{17}\end{bmatrix}
=
\tfrac{1}{\sqrt{31}} \left(1\cdot \ket{00} + 2 \cdot \ket{01} -3 \cdot \ket{10} + \sqrt{17} \ket{11}\right)
\]
\[
\ket{\psi} \defn
\frac{1}{2} \cdot \begin{bmatrix} 1\\ -1\\ 1\\ -1\end{bmatrix}
,\qquad\quad
\ket{\eta} \defn 
\tfrac{1}{\sqrt{2}} \cdot \begin{bmatrix} 1, 0, 0, 0, 0, 0, 0, 1\end{bmatrix}^{\top} = \tfrac{1}{\sqrt{2}} \ket{000} + \tfrac{1}{\sqrt{2}} \ket{111}
.
\]

\vspace{-2.em}
\hfill$\diamond$
\end{example}
\vspace{-1em}

Consider a set of six qubits and suppose that the first two are jointly in some arbitrary state $\ket{\phi}$ (a vector of length $2^2 = 4$) and the remaining four in some state $\ket{\psi}$ (a vector of length $2^4 = 16$).
Identically to the case of probabilistic computing (\autoref{sec:probabilistic-computing}), we use the Kronecker product $\otimes$ to find the joint state on the six qubits, which is $\ket{\phi} \otimes \ket{\psi}$ (a vector of length $2^2 \cdot 2^4 = 2^6$).
\exercise{Show that the computational-basis states $\ket{b}$ for $b=(b_1,b_2,\dots,b_n) \in \{0, 1\}^n$ can be written as $\ket{b} = \ket{b_1} \otimes \ket{b_2} \otimes \dots \otimes \ket{b_n}$.}

\subsubsection{Manipulating quantum states using gates.}
\label{sect:quantum-info--gates}

A quantum gate on $n$ qubits is represented by a $2^n$ by $2^n$ unitary matrix $U$ (unitarity defined below).
A quantum state $\ket{\phi}$ is updated by a unitary matrix as $U\cdot \ket{\phi}$ where $\cdot$ denotes matrix-vector multiplication.

Recall from \autoref{sec:background-intro} that the three-(qu)bit $CCNOT$ gate (or doubly control-NOT, or Toffoli gate) is universal for classical computing, while the $NOT$ and $CNOT$ gates can be added as syntactic sugar.
Below we give their matrices.
To obtain a universal gate set for quantum computing, we merely need to extend this gate set with the single-qubit \emph{Hadamard gate}: $H\, \defn
    \frac{1}{\sqrt{2}} \left[\begin{smallmatrix*}[r]
      1 & 1 \\
      1 & -1
    \end{smallmatrix*}\right].$ 
    \autoref{ex:h} illustrates its function: realizing superpositions.
\[NOT \defn 
  \left[\begin{matrix}
    0 & 1 \\
    1 & 0 \\    
  \end{matrix}\right]\hspace{-1mm}
  , 
  ~~
  CNOT \defn 
  \left[\begin{smallmatrix}
    1 & 0 & 0 & 0 \\
    0 & 1 & 0 & 0 \\
    0 & 0 & 0 & 1 \\
    0 & 0 & 1 & 0 \\    
  \end{smallmatrix}\right]\hspace{-1mm}
  , 
  ~~ 
  CCNOT \defn \left[\begin{smallmatrix}
    1 & 0 & 0 & 0 & 0 & 0 & 0 & 0 \\
    0 & 1 & 0 & 0 & 0 & 0 & 0 & 0 \\
    0 & 0 & 1 & 0 & 0 & 0 & 0 & 0 \\
    0 & 0 & 0 & 1 & 0 & 0 & 0 & 0 \\
    0 & 0 & 0 & 0 & 1 & 0 & 0 & 0 \\
    0 & 0 & 0 & 0 & 0 & 1 & 0 & 0 \\
    0 & 0 & 0 & 0 & 0 & 0 & 0 & 1 \\
    0 & 0 & 0 & 0 & 0 & 0 & 1 & 0 \\
  \end{smallmatrix}\right]\hspace{-1mm}
\]
\begin{example}\label{ex:h}
    Applying the $H$ gate to the state $\ket{0}$ or $\ket{1}$, we obtain
\begin{align*}
&       H\cdot\ket{0} = 
   \frac{1}{\sqrt{2}}\begin{bmatrix*}[r]
    1 & 1 \\
    1 & -1
    \end{bmatrix*} \cdot 
    \begin{bmatrix}
    1 \\
    0
    \end{bmatrix} = 
    \begin{bmatrix}
    \frac{1}{\sqrt{2}} \\
    \frac{1}{\sqrt{2}}
    \end{bmatrix} = \frac{1}{\sqrt{2}}(\ket{0}+\ket{1}),
\\
&   H\cdot\ket{1} = 
   \frac{1}{\sqrt{2}}\begin{bmatrix*}[r]
    1 & 1 \\
    1 & -1
    \end{bmatrix*} \cdot 
    \begin{bmatrix}
    0 \\
    1
    \end{bmatrix} = 
    \begin{bmatrix}
    \frac{1}{\sqrt{2}} \\
    -\frac{1}{\sqrt{2}}
    \end{bmatrix} = \frac{1}{\sqrt{2}}(\ket{0}-\ket{1})
.
\end{align*}
 
    \vspace{-2em}
    \hfill$\diamond$
\end{example}

A square matrix $U$ is unitary if it is invertible and its inverse is $U^{\dagger}$, where $U^{\dagger}$ is found by transposing $U$ for real matrices.\footnote{The adjoint $U^{\dagger}$ of a complex matrix $U$ is found by transposing $U$, followed by replacing each matrix entry $u + wi$ with its complex conjugate $u - wi$. As we use only real matrices in this paper, one can always think of the adjoint as the transpose. The adjoint notation $U^{\dagger}$ is used in this paper, in favor of the inverse $U^{-1}$, as this is common in the quantum community.}
Unitarity ensures that the matrix is norm-preserving, thus guaranteeing that the output quantum state has norm 1 if the input state does so too.\footnote{Quantum states should have norm 1 to be physically meaningful since the probabilites over the computational-basis states should sum up to 1. Further, quantum gates need to be unitary to obey the energy conservation law, due to quantum mechanical properties following from the famous Schödinger equation.}
\newl{As unitary matrices are reversible, we directly see that all quantum gates are also reversible. This relates quantum computing to reversible computing.}
The Hadamard gate and NOT gate have adjoint operators $H^{\dagger} = \frac{1}{\sqrt{2}}
\left[\begin{smallmatrix}
    1 & 1 \\
    1 & -1
\end{smallmatrix}\right]
=H
$ and
$
NOT^{\dagger} = 
\left[\begin{smallmatrix}
    0 & 1 \\
    1 & 0
\end{smallmatrix}\right]
= NOT$
; it is not hard to check, indeed, that $H^{\dagger} \cdot H = NOT^{\dagger} \cdot NOT = \unit = \left[\begin{smallmatrix} 1 & 0 \\ 0 & 1 \end{smallmatrix}\right]$.

To apply a single-qubit gate to one qubit of a state on $n>1$ qubits, 
one uses the Kronecker product (see \autoref{sec:background-intro}) for `padding' the gate with an identity matrix $I$ for each other qubit, i.e. the matrices which leave any vector unchanged.
For example, applying $H$ to the first qubit of $\ket{010}$ is computed as $(H\otimes I \otimes I)\ket{010}$.

\exercise{Compute the unitary matrix $H\otimes I \otimes I$.
Remark that it has dimensions $2^3 \times 2^3$, as is needed for being able to apply it to a $3$-qubit state.}

Finally, we remark that sequential composition of gates is represented by matrix multiplication of their unitaries, similar to matrix multiplication in the Markov chain case in \autoref{sec:background-intro}: applying first $U$ and then $V$ to $\ket{\phi}$ yields the state $V \cdot (U \cdot \ket{\phi}) = (V\cdot U) \ket{\phi}$.

\subsubsection{Measurement}
\label{subsec:measurement}
Measurement enables one to extract classical information (bits) from a quantum state.
The theory of quantum measurement allows many ways to do this and here we only focus on a common one: measuring all qubits in the computational basis.
Given an $n$-qubit quantum state $\sum_{b\in \{0, 1\}^n} \alpha_b \ket{b}$,
a measurement (on all qubits) is a probabilistic operation that returns one of the values $b\in\{0,1\}^n$, called the \concept{measurement outcome}.
The value $b$ is returned with probability $(\alpha_{b})^2$.
(Or $|\alpha_{b}|^2$ in the case of complex amplitudes.)    
For example, the probability of measuring $00\dots 0$ is $(\alpha_{00\dots0})^2$.
Since each quantum state vector has unit norm, these probabilities add up to 1, as desired.

\begin{example}\label{ex:measurement}
Consider the state $\begin{bmatrix}
    \frac{1}{2},
    0,
    \frac{1}{2},
    0,
    \frac{1}{2},
    0,
    0,
    \frac{1}{2}
\end{bmatrix}^{\top} = \frac{1}{2}\ket{000}+\frac{1}{2}\ket{010}+\frac{1}{2}\ket{100}+\frac{1}{2}\ket{111}$.
Upon measurement, the probability to obtain measurement outcome 000 is $(\alpha_{000})^2 = \left(\frac{1}{2}\right)^2 = \frac{1}{4}$ and the probability to obtain measurement outcome 001 is $(\alpha_{001})^2=0^2=0$.
\hfill$\diamond$

\end{example}

\subsubsection{Quantum circuit}
\label{prelim:quantumcircuits}
A quantum circuit is composed of qubits represented by horizontal lines (wires) and quantum gates represented by boxes, with each gate acting on one or more qubits.
The circuit is finished with a measurement.
We will always let the input state be $\ket{0}\otimes\ket{0}\otimes\dots\otimes\ket{0}$ (which is usually written as $\ket{0}^{\otimes n}$ or $\ket{00\dots0}$), and the output state is obtained after the sequential application of the circuit's gates.
All gates can be combined with matrix multiplication into a single unitary operator describing the entire circuit.

{
\begin{example}
\label{example:easy-circuit}
In the following circuit, we apply a Hadamard to the first two qubits of $\ket{000}$. Then we apply a Toffoli gate ($CCNOT$) to the three qubits, followed by an all-qubit measurement. The intermediate states are as follows; see \autoref{ex:measurement} for evaluating measurement.

\label{ex:bell-state}
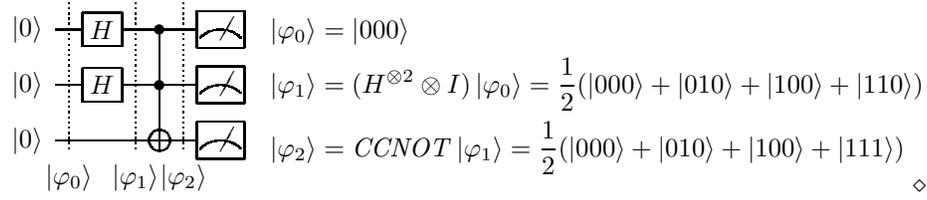
\begin{minipage}{0.27\textwidth}
\hspace*{0.5cm}
$
    \begin{array}{c}  
      \Qcircuit @C=.5em @R=.7em {
        \lstick{\ket{0}} & \qw\ar@{.}[]+<0em,1em>;[d]+<0em,-2.4em> & \gate{H} &\qw\ar@{.}[]+<0em,1em>;[d]+<0em,-2.4em> & \ctrl{2} & \qw\ar@{.}[]+<0em,1em>;[d]+<0em,-2.4em> & \meter \\
        \lstick{\ket{0}} & \qw & \gate{H} & \qw & \ctrl{1} & \qw  &\meter\\
        \lstick{\ket{0}} & \qw & \qw & \qw & \targ & \qw &\meter \\
        & \ket{\varphi_0} & & \ket{\varphi_1} & & \ket{\varphi_2} & & 
      }
  \end{array}
$
\end{minipage}
\begin{minipage}{0.729\textwidth}
\vspace{-1em}
\begin{align*}
    \ket{\varphi_0} &= \ket{000}\\
    \ket{\varphi_1} &= (H^{\otimes2}\otimes I) \ket{\varphi_0} = \frac{1}{2}(\ket{000}+\ket{010}+\ket{100}+\ket{110})\\
    \ket{\varphi_2} &= \mathit{CCNOT}\ket{\varphi_1} = \frac{1}{2}(\ket{000}+\ket{010}+\ket{100}+\ket{111})
\end{align*}
\end{minipage}
\vspace{-.8em}
\hfill$\diamond$
\end{example}
}

We emphasize that quantum circuits are reversible: each gate, and hence each circuit, has the same number of input and output qubits.

Suppose that we have an $n$-qubit circuit $C$ initialized to the all-zero state $\ket{00\dots0}$.
Computing the probability of outcome $b\in\{0,1\}^n$ can now be written as follows:
$|\bra{b} C \ket{00\dots 0}|^2$. Here, $\bra{b}$ is a row vector of the computational basis state $\ket b$, so $\bra{b} C \ket{00\dots 0}$ can be seen as a product of a row vector, matrix and column vector, resulting in a scalar. 

\begin{example}
    Let $C$ be the circuit from \autoref{example:easy-circuit}. Then $C\ket{000}$ equals $\frac{1}{2}\ket{000}+\frac{1}{2}\ket{010}+\frac{1}{2}\ket{100}+\frac{1}{2}\ket{111}= \begin{bmatrix}
    \frac{1}{2},
    0,
    \frac{1}{2},
    0,
    \frac{1}{2},
    0,
    0,
    \frac{1}{2}
    \end{bmatrix}^{\top}$.
    Hence, the probability of measuring $000$ is $|\bra{000}C\ket{000}|^2 = \frac{1}{4}$ and the probability of measuring $001$ is $|\bra{001}C\ket{000}|^2 = 0$.

    Note that, as in \autoref{sec:probabilistic-computing}, we use a row vector for the measured computational basis state and a column vector for the initial state of the circuit.
\end{example}

\keymessage{an $n$-qubit state is transformed by $2^n \times 2^n$ matrices through matrix-vector multiplication. Measurement allows one to extract information from the state.
}

We will now take a first step towards using quantum circuits to our advantage.
We focus on Boolean satisfiability (SAT), and will provide a naive approach to solving SAT using a quantum computer.
Although this approach will not work, it will show how the \emph{single} evaluation of the Boolean circuit on a quantum state will evaluate the Boolean function on all \emph{exponentially-many} bitstrings as input.

Consider the 3-CNF formula $f(x,y,z) = x \wedge (\neg x\vee y)\wedge(\neg x\vee\neg y\vee z)$. We will use a reversible circuit to implement this function as a quantum circuit. We usually call such quantum circuit a \textit{function oracle}. 
We want to compute $f$ for every possible input $000,001,\dots,111$. Therefore, we create a superposition of these states. This is done by applying a Hadamard gate to the qubits representing the input (see exercise below).

Now we will see what happens if we apply the function oracle $f$ to a superposition of input states. \autoref{example:easy-circuit} contains a simple example of a quantum circuit calculating $f=x\wedge y$. A more complicated example is the following one:

{
\begin{example}
\label{ex:quantum-sat}
Consider the 3-CNF formula
\begin{equation}
\label{eq:three-cnf}
f(x,y,z) = x \wedge (\neg x\vee y)\wedge(\neg x\vee\neg y\vee z).
\end{equation}
The following circuit first applies Hadamard gates to get a superposition on the input qubits. Then it applies the function oracle for $f$ (in the dashed rectangle).
$$
    \hspace*{-3em}\begin{array}{c}  
      
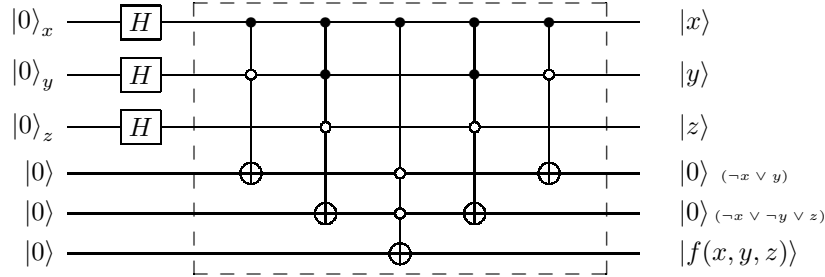
\Qcircuit @C=1em @R=.7em {
        \lstick{\ket{0}_x} & \qw & \gate{H} &\qw  & \qw &\ctrl{1} & \qw & \ctrl{1} & \qw  & \ctrl{3}& \qw & \ctrl{1} & \qw &\ctrl{1} & \qw & \qw& \qw & \rstick{\ket{x}}\\
        \lstick{\ket{0}_y} & \qw & \gate{H} & \qw & \qw & \ctrlo{2} &\qw&\ctrl{1} & \qw & \qw & \qw & \ctrl{1} & \qw &  \ctrlo{2} & \qw  & \qw& \qw & \rstick{\ket{y}}\\
        \lstick{\ket{0}_z} & \qw & \gate{H} &  \qw & \qw &\qw & \qw & \ctrlo{2} & \qw& \qw & \qw & \ctrlo{2} & \qw& \qw& \qw& \qw& \qw & \rstick{\ket{z}}\\
        \lstick{\ket{0}} & \qw & \qw &  \qw & \qw &\targ & \qw & \qw  & \qw & \ctrlo{1} & \qw& \qw& \qw&\targ& \qw& \qw& \qw & \rstick{\ket{0} \text{\tiny{ ($\neg x \vee y)$}}}\\
        \lstick{\ket{0}} & \qw &  \qw & \qw & \qw &\qw & \qw & \targ& \qw&  \ctrlo{1} & \qw & \targ & \qw&  \qw& \qw& \qw& \qw& \rstick{\ket{0} 
        \text{\tiny{$(\neg x\vee \neg y \vee z)$}  }   } \\
        \lstick{\ket{0}} & \qw & \qw & \qw & \qw &\qw & \qw & \qw & \qw & \targ & \qw& \qw& \qw& \qw& \qw& \qw& \qw &\rstick{\ket{f(x,y,z)}} \gategroup{1}{5}{6}{15}{1.5em}{--}\\
      }
  \end{array}
$$

The resulting state is $\sum_{x,y,z\in\{0,1\}^3}\ket{x,y,z}\otimes\ket{f(x,y,z)}$, or, written out in full, 
\begin{equation}
\label{eq:cnf-result}
\frac{1}{2\sqrt{2}}(\ket{0000}+\ket{0010}+\ket{0100}+\ket{0110}+\ket{1000}+\ket{1010}+\ket{1100}+\ket{1111})
.
\end{equation}
(Here, we omitted the two auxiliary qubits uncomputed to $\ket 0$). Note that the single satisfying assignment (111) represents the solution to the satisfiability problem for $f$, because in \autoref{eq:cnf-result}, the only term which has a `$1$' at the fourth qubit, corresponding to $f(x,y,z)$, is $\ket{1111}$. 

\exercise{Evaluate the circuit above in three steps. \textbf{(a)} Show that applying $H\otimes H \otimes H$ to the input $\ket{000}$ (the top three qubits in the circuit) yields a superposition over all $3$-qubit computational-basis states. \textbf{(b)} Next, find the state when adding the register of the bottom three qubits (which is in the state $\ket{000}$) using the Kronecker product. 
\textbf{(c)} Lastly, using the Boolean function $f$ as given in \autoref{eq:three-cnf} (the circuit in the dashed rectangle above), verify that the resulting state indeed is \autoref{eq:cnf-result}.}

We would like to extract the satisfying assignment `111' using measurement. Unfortunately, we see that measuring gives outcome 1111 with a probability of only $\left|\frac{1}{2\sqrt{2}}\right|^2 = \frac{1}{8}$. This probability is $\frac{1}{2^n}$ in general, where $n$ is the number of variables to $f$. 
Thus, finding a satisfiable instance this way only succeeds with exponentially-small probability, much like the classical naive approach of randomly guessing bitstrings as input and evaluating $f$ on them.
\hfill$\diamond$
\end{example}

\explainerbox{Boolean satisfiability \& quantum computing.}{
In \autoref{ex:quantum-sat} above, we saw a naive approach to solving Boolean satisfiability using quantum computing, which performed equally well as random guessing.
However, using the famous quantum algorithm by Grover~\cite{grover1996fast}, the probability of measuring a satisfying assignment of $f$ can be increased to arbitrarily high probability.
Since every CNF formula $f$ can be efficiently encoded as a reversible circuit,
this solves the satisfiability problem for $f$.
The Grover algorithm has runtime $O^*(\sqrt{2^n})$, where $O^*$ omits polynomial factors. This is a quadratic speed-up compared to brute force and the best-known classical algorithm for $k$-SAT where $k$ is unbounded.\footnote{Sch\"oning's~\cite{schoning1999probabilistic} and the PPSZ~\cite{paturi2005improved} algorithm deliver better guarantees for constant $k$, e.g., for 3-SAT, and can also be quantized~\cite{Rennela2023hybriddivideconquer}.} 
}
}

\subsection{Visualizing a quantum computation using an automaton}
\label{sec:visualizing}

Armed with a well-defined notion of a quantum circuit, we can now continue the comparison with Markov chains in \autoref{sec:background-intro} and visualize an $n$-qubit quantum gate as an automaton: it has $2^n$ states, one for each computational-basis state, and we interpret the matrix of each gate $U$ as the weighted adjacency matrix between these states.
For example, consider $2$ qubits, acted upon by the $CNOT$ gate and by the $H$ gate on the first qubit:

\begin{minipage}{.5\textwidth}
$
\textcolor{red}{H \otimes I = \left[\begin{smallmatrix} \nicefrac{1}{\sqrt{2}} & 0 & \nicefrac{1}{\sqrt{2}} & 0\\ 0 & \nicefrac{1}{\sqrt{2}} & 0 & \nicefrac{1}{\sqrt{2}}\\ \nicefrac{1}{\sqrt{2}} & 0 & -\nicefrac{1}{\sqrt{2}} & 0 \\ 0 & \nicefrac{1}{\sqrt{2}} & 0 & -\nicefrac{1}{\sqrt{2}}\end{smallmatrix}\right]
}
,\quad
\textcolor{blue}{
CNOT = \left[\begin{smallmatrix} 1 & 0 & 0 & 0\\ 0 & 1 & 0 & 0\\ 0 & 0 & 0 & 1\\ 0 & 0 & 1 & 0\end{smallmatrix}\right]
}
$
\end{minipage}
\hspace{1cm}
\begin{minipage}{.5\textwidth}
\begin{center}
\scalebox{0.7}{\begin{tikzpicture}[
    scale=0.3,
    every path/.style={>=latex},
    every node/.style={},
    inner sep=0pt,
    minimum size=14pt,
    line width=1pt,
    node distance=.5cm,
    thick,
    font=\footnotesize
    ]

    \node[draw, circle, minimum size=0.9cm] (00)   {$\ket{00}$};
    \node[draw,circle, right =1cm of 00, xshift=0cm, minimum size=0.9cm] (10) {$\ket{10}$};
    \node[draw,circle, below =1cm of 00, xshift=0cm, minimum size=0.9cm] (01) {$\ket{01}$};
    \node[draw,circle, below =1cm of 10, xshift=0cm, minimum size=0.9cm] (11) {$\ket{11}$};

\draw[red, \bigarrowhead, bend left=20] (00) edge node[yshift=0.2cm] {$\nicefrac{1}{\sqrt{2}}$} (10);
\draw[loop left, red, \bigarrowhead] (00) edge node[yshift=0.2cm, out=45, in=90] {$\nicefrac{1}{\sqrt{2}}$} (00);
\draw[loop above, blue, \bigarrowhead] (00) edge node[yshift=0.0cm] {$1$} (00);

\draw[red, \bigarrowhead, bend left=20] (10) edge node[yshift=-0.2cm] {$\nicefrac{1}{\sqrt{2}}$} (00);
\draw[loop right, red, \bigarrowhead] (10) edge node[yshift=0.2cm, out=45, in=90] {$-\nicefrac{1}{\sqrt{2}}$} (10);
\draw[bend right=20, blue, \bigarrowhead] (10) edge node[xshift=-0.2cm, out=45, in=90] {$1$} (11);

\draw[red, \bigarrowhead, bend left=20] (01) edge node[yshift=0.2cm] {$\nicefrac{1}{\sqrt{2}}$} (11);
\draw[loop left, red, \bigarrowhead] (01) edge node[yshift=0.2cm, out=45, in=90] {$\nicefrac{1}{\sqrt{2}}$} (01);
\draw[loop below, blue, \bigarrowhead] (01) edge node[yshift=0.0cm] {$1$} (01);

\draw[red, \bigarrowhead, bend left=20] (11) edge node[yshift=-0.2cm] {$\nicefrac{1}{\sqrt{2}}$} (01);
\draw[loop right, red, \bigarrowhead] (11) edge node[yshift=0.2cm, out=45, in=90] {$-\nicefrac{1}{\sqrt{2}}$} (11);
\draw[bend right=20, blue, \bigarrowhead] (11) edge node[xshift=0.2cm, out=45, in=90] {$1$} (10);

\end{tikzpicture}}
\end{center}
\end{minipage}
We will now use the following known result from graph theory:
\begin{quote}
The entry in the product of adjacency matrices at row $r$ and column $c$ equals the sum of products of edge labels (a \emph{path sum}) over all paths from node $r$ to node $c$.
\end{quote}
This somewhat complicated statement tells us that matrix multiplication can be visualized as path sums, a result we already implicitly used in the Markov chain case in \autoref{sec:probabilistic-computing} when observing that sequential composition of gates is represented by multiplication of transition matrices.
To illustrate how this statement results in a visualization of quantum gates as paths in the automaton, we give the following example.
\begin{example}
Consider the circuit below; we will compute the entries in $(H\otimes I) \cdot CNOT\cdot CNOT \cdot (H\otimes I)$ corresponding to the transitions $\ket{00} \rightarrow \ket{00}$ and $\ket{00} \rightarrow \ket{10}$, first using matrix-vector multiplication, and then using paths in the automaton.
For the former, it is straightforward to derive that $\ket{\varphi_1} = \stwo (\ket{00} + \ket{10})$, hence $\ket{\varphi_3} = \stwo \cdot 1 \cdot 1 \cdot \ket{00} + \stwo \cdot 1 \cdot 1 \cdot \ket{10}$ (the second CNOT uncomputes the first). We use this to compute $\ket{\varphi_4} = (H\otimes I)\ket{\varphi_3}$ below.
\newline
\begin{minipage}{0.33\textwidth}
\hspace{.5cm}
\scalebox{0.8}{$
    \begin{array}{c}  
      \Qcircuit @C=-.33em @R=.7em @!C @!R{
        \lstick{\ket{0}} & \qw\ar@{.}[]+<0em,1em>;[d]+<0em,-0.5em> & \gate{\textcolor{red}{H}} &\qw\ar@{.}[]+<0em,1em>;[d]+<0em,-0.5em> & \ctrl{1} & \qw\ar@{.}[]+<0em,1em>;[d]+<0em,-0.5em> & \ctrl{1} & \qw\ar@{.}[]+<0em,1em>;[d]+<0em,-0.5em> & \gate{\textcolor{red}{H}} & \qw\ar@{.}[]+<0em,1em>;[d]+<0em,-0.5em> & \qw \\
        \lstick{\ket{0}} & \qw & \qw & \qw & \qw{~~\textcolor{blue}{\oplus}~~} & \qw &  \qw{\textcolor{blue}{\oplus}} & \qw & \qw & \qw & \qw \\
        & \ket{\varphi_0} & & \ket{\varphi_1} & & \ket{\varphi_2} & & \ket{\varphi_3} & & \ket{\varphi_4}
      }
  \end{array}
$}
\end{minipage}
\begin{minipage}{0.66\textwidth}
{\small\begin{eqnarray}
\ket{\varphi_4}
\nonumber
&=& \tfrac{1}{\sqrt{2}} \cdot 1 \cdot 1 \cdot (\tfrac{1}{\sqrt{2}}  \ket{00} + \tfrac{1}{\sqrt{2}}  \ket{10} +  \tfrac{1}{\sqrt{2}}  \ket{00} - \tfrac{1}{\sqrt{2}}  \ket{10})
\nonumber
\\
&=&
\left(
\tstwo \cdot 1 \cdot 1 \cdot \tstwo +
\tstwo \cdot 1 \cdot 1 \cdot \tstwo
\right)
\ket{00}
\nonumber
\\&&\quad
+
\left(
\tstwo \cdot 1 \cdot 1 \cdot \tstwo -
\tstwo \cdot 1 \cdot 1 \cdot \tstwo
\right)
\ket{10}
\label{eq:interference}
\\
&=&
\left(
\tfrac{1}{2}
+\tfrac{1}{2}
\right)
\ket{00}
+
\left(
\tfrac{1}{2}
-\tfrac{1}{2}
\right)
\ket{10}
\label{eq:interference-continued}
\end{eqnarray}}
\end{minipage}
\vspace{0cm}
\newline
Now the gate sequence $H-CNOT-CNOT-H$ means we should consider `red-blue-blue-red' paths in the automaton.
The factor $\frac{1}{2} + \frac{1}{2}$ in front of $\ket{00}$ in \autoref{eq:interference-continued} is mirrored in the automaton by noting that there are two paths from $\ket{00}$ to itself:
{\circled{$\ket{00}$}
$\textcolor{red}{\stackrel{\nicefrac1{\sqrt{2}}}{\rightarrow}}$
\circled{$\ket{10}$}
$\textcolor{blue}{\stackrel{1}{\rightarrow}}$
\circled{$\ket{11}$}
$\textcolor{blue}{\stackrel{1}{\rightarrow}}$
\circled{$\ket{10}$}
$\textcolor{red}{\stackrel{\nicefrac1{\sqrt{2}}}{\rightarrow}}$
\circled{$\ket{00}$}}
and
{\circled{$\ket{00}$}
$\textcolor{red}{\stackrel{\nicefrac1{\sqrt{2}}}{\rightarrow}}$
\circled{$\ket{00}$}
$\textcolor{blue}{\stackrel{1}{\rightarrow}}$
\circled{$\ket{00}$}
$\textcolor{blue}{\stackrel{1}{\rightarrow}}$
\circled{$\ket{00}$}
$\textcolor{red}{\stackrel{\nicefrac1{\sqrt{2}}}{\rightarrow}}$
$\circled{$\ket{00}$}.$}
These paths both have amplitude $\nicefrac{1}{\sqrt{2}} \cdot 1 \cdot 1 \cdot \nicefrac{1}{\sqrt{2}} = \nicefrac{1}{2}$, and their sum $\nicefrac{1}{2} + \nicefrac{1}{2}$ is the amplitude of $\ket{00}$ in $\ket{\varphi_4}$ in \autoref{eq:interference-continued} indeed (also note that the two products $\nicefrac{1}{2} \cdot 1 \cdot 1 \cdot \nicefrac{1}{2}$ are precisely the terms in front of $\ket{00}$ in \autoref{eq:interference}).
In contrast, the factor $\nicefrac{1}{2} - \nicefrac{1}{2}$ in front of $\ket{10}$ in \autoref{eq:interference} arises in the automaton as two paths from $\ket{00}$ to $\ket{10}$, one with amplitude $\nicefrac{1}{\sqrt{2}} \cdot 1 \cdot 1 \cdot \nicefrac{1}{\sqrt{2}}$, and one with amplitude $\nicefrac{1}{\sqrt{2}} \cdot 1 \cdot 1 \cdot -\nicefrac{1}{\sqrt{2}}$, which are precisely the terms in front of $\ket{10}$ in \autoref{eq:interference}.
\hfill$\diamond$\end{example}
\vspace{-1em}
Note that in the example above, the path sums have opposite sign, so they precisely cancel each other, implying that the transition $\ket{00}\rightarrow \ket{10}$ has amplitude zero!
This cannot happen in a Markov chain as probabilities are never negative.

We thus see that the transition amplitude from state $\ket{x}$ to state $\ket{y}$, which we first found through linear algebra, can be found in the automaton by summing path contributions from \circled{$\ket{x}$} to \circled{$\ket{y}$}, where each path contribution is the product of the edge labels of the path.

The automaton is useful because it shows a few properties of quantum circuits:
\begin{enumerate}
\item an \textbf{exponentially-sized state space} as function of number of qubits
\item the \textbf{combinatorial nature of the evolution of a quantum state through a circuit:} there can be \textbf{many paths} from one node to another, sometimes exponentially-many, and we need to track all of them to compute the output state's amplitudes
\item the many paths from one node to another will \textbf{interfere} as amplitudes, either \textbf{constructively} (the amplitudes amplify through addition, as in the example above for computing the amplitude of $\ket{00}$) or \textbf{destructively} (the amplitudes cancel, as for $\ket{10}$ above).
\end{enumerate}
Item 1 and 2 also occur for probabilistic computation (see the Markov chain in \autoref{sec:background-intro}) but item 3 is where quantum computing differs.
Automated reasoning methods were often developed to solve scenarios where items 1 and 2 are present; we will see one approach in \autoref{sec:sat} that also illustrates item 3.

\keymessage{during a run of a quantum circuit, there can be exponentially many paths leading from one state to another.
The amplitude contributions of these paths can constructively or destructively interfere.
}

\subsection{Towards a quantum advantage}

Application of a gate $G$ once to a superposition of many computational-basis states $\ket{b}$ yields a sum of terms of the form $G\ket{b}$.
This realization is particularly astounding in case we start out with an exponentially-large superposition, which can be created already with only linearly many gates in the number of qubits (recall the exercise above to compute $H^{\otimes n} \ket{0}^{\otimes n}$ for $n=3$).
However, using measurement, we can only read off a single bitstring from this superposition.

A plethora of quantum algorithms\footnote{A quantum algorithm is a uniform family of quantum circuits (like in circuit complexity~\cite{arora2009computational}).}~\cite{quantumalgorithmzoo} has been found whose quantum circuits provably need polynomially-fewer or exponentially-fewer gates than their classical counterpart.\footnote{The exact statements depend on the used model (for example, whether the input is promised to be picked from a certain set). Quantum algorithms are typically complex, so we decided they are out of scope for this tutorial, where we aim to focus on analyzing quantum circuits using formal methods tools.}
Real-world quantum devices suffer from noise, and counteracting that noise might require additional resources that cancel the complexity-theoretic and real-world advantages of quantum computing.
Arguably \textbf{the main open problem of quantum technologies is therefore to provide a real-world demonstration of a quantum advantage}.
On the road to this goal, there are several tasks where formal methods could help: 
performance prediction of real-world devices through classical simulation of quantum circuits (sampling or computing the measurement distribution);
optimizing circuits (e.g. fewer gates, circuit layouts following the topology of real-world chips, etc.);
verification, specifically checking if two quantum circuits implement the same unitary matrix;
finding a circuit that outputs a desired quantum state;
transpilation to a gate set that real-world devices can natively run;
etc.

In the remainder of this tutorial, we will focus on using
weighted model counting or \#SAT (\autoref{sec:sat}) 
to perform classical simulation of quantum circuits.

\subsection{Our scope: Quantum circuit simulation}
\label{sec:simulation}

In this tutorial, we will mainly consider the task of \emph{classically simulating a quantum circuit} (that is, to design classical algorithms for this task).
Formally, the task of simulating an $n$-qubit quantum circuit $C$ is to find the probability of outcome $b\in\{0,1\}^n$ when the output state of $C$ is measured, assuming that $\ket{0}^{\otimes n}$ is the input quantum state to $C$.
Although simulation is \#P-hard in general~\cite{nest2008classical}, so are many problems in formal methods, where solutions have been found that work well in practice.
We should thus not be discouraged from tackling simulation and will see two approaches to do so in \autoref{sec:sat} and \autoref{sec:dd}. In \autoref{sec:outro}, we refer to extensions that tackle other important quantum circuit analysis tasks.

We focus here on \emph{strong simulation}~\cite{Chen:2023rwy}: the problem of returning a probability for a certain computational basis state. In contrast, \emph{weak simulation} asks to sample the probability distribution of measurement outcomes (and hence is inherently probabilistic).

\section{Reducing Quantum Computing to \#SAT}
\label{sec:sat}

Here we reduce quantum-circuit simulation to weighted model counting (weighted-\#SAT).
This section is partly based on \cite{mei2024simulating}, while the approach here effectively realizes the well-known path-sum approach~\cite{feynman2010quantum}.
An extended version of this encoding is given in \aref{app:pauli}.

\subsection{SAT and \#SAT}
We denote $SAT(F):= \set{\alpha \mid F(\alpha) = 1}$ for the set of all satisfiable assignments of a propositional formula $F \colon \{0, 1\}^V \to \{0, 1\}$ over a finite set of Boolean variables $V$.
We say that $F$ is \emph{satisfiable} if $SAT(F)$ is non-empty.
We write an assignment $\alpha$ as a \concept{cube} (a conjunction of literals, i.e., positive or negative variables), e.g., $a\land b$, or shorter~$ab$.

\begin{wrapfigure}[4]{r}{5cm}
\vspace{-2.4em}
\begin{circuitikz}
\draw
(0,2) node[and port] (myand) {\hspace{-1ex}$\land$}
(2,2) node[not port] (mynot) {\hspace{-1ex}$\neg$}
(myand.in 1) node[anchor=east] {$a$}
(myand.in 2) node[anchor=east] (bnode) {$b$}
(myand.out) -- (mynot.in 1)
(mynot.out) node[anchor=west] {$c$}
;
\end{circuitikz}
\end{wrapfigure}

The action of a classical circuit can be encoded by SAT constraints directly by representing each bit as a Boolean variable. For example,
the Boolean constraint for the classical circuit $C$ on the right is $ F_C(V) = c \Leftrightarrow \neg (a\wedge b)$ over variables $V=\{a,b,c\}$.
  Given an input $a=0$ and $b=1$,
  the satisfying assignment is $\alpha = \no a bc$,
  where $\alpha(c) = 1$ is the final state.

We denote $\#SAT(F) \defn |SAT(F)|$ for the \emph{model count} of a formula $F$.
A weight function $W\colon \set{ \no v, v \mid v\in V} \to \mathbb{R} $ assigns 
a real-valued weight to positive literals $v$ (i.e., $v = \true$) and the negative literals $\no v$ (i.e., $v = \false$).
We say variable $v$ is \emph{unbiased} iff $W(v) = W(\no v) = 1$.
Given an assignment $\alpha\in\bool^V$,
let $W(\alpha(v)) = W(v := \alpha(v))$ for $v\in V$.
For a propositional formula $F$ over $V$ and a weight function $W$, 
we define \emph{weighted model counting} (\#SAT) as:
\[
\#SAT_W(F) \defn   \sum_{\alpha \in  SAT(F) }  W(\alpha)\text{ where }  W(\alpha) = \prod_{v\in V} W(\alpha(v)).
\]

\begin{example}
The propositional formula $F = (v_1 \vee v_2) \wedge (\no{v_1} \vee v_2) \wedge v_3$ over $V=\{v_1,v_2, v_3\}$ has two satisfying assignments: $\alpha_1 = v_1  v_2  v_3$ and $\alpha_2 = \no{v_1}  v_2  v_3$.
  We define the weight function $W$ as $W(v_1) = -\frac{1}{2}$, $W(\no{v_1}) = \frac{1}{3}$ and $W(v_2) = \frac14$, $W(\no{v_2}) = \frac34$, while $v_3$ remains unbiased.
  The weight of $F$ can be computed as $MC_W(F) = -\frac{1}{2} \times \frac14  \times 1 + \frac{1}{3}  \times \frac14  \times 1  = -\frac{1}{24}$.
\end{example}

\explainerbox{Why \#SAT for tackling simulation?}{
First, analysis tasks on quantum circuits are inherently functional problems, as the outcome often is a measurement probability (or amplitude). For the same reason, inference on Bayesian Networks~\cite{roth1996hardness,sang2005performing} was done with weighted model counting. Here we use a similar approach by reducing the simulation of quantum circuits to weighted model counting.

In our \#SAT encoding, we let each satisfying assignment encode 
a path in the automaton as discussed in \autoref{sec:visualizing}.
We then add weighted variables so that the weight for each assignment (representing a path) equals a part of the circuit's final amplitude. We also let the measurements constrain the encoding so that only the required paths remain. The weighted model counter ensures that the weights of all paths contributing to the measurement outcome are summed up (positive and negative).

Our encoding uses only linearly many variables and clauses in the number of gates plus the number of qubits. So, to encode a single gate, we require only a constant amount of clauses (around four), while encoding measurements require $n$ (unit) clauses, one per qubit.
}

\subsection{Encoding Quantum States using (Weighted) Boolean Variables}\label{sec:encoding-state}

Recall from \autoref{sec:prelims} that a quantum state $\ket{\phi}$ is a specific linear combination of classical states, i.e., 
\[\ket{\phi} = \sum_{b \in \set{0,1}^n} \alpha_{b} \ket{b}.\]
Accordingly, we can simply reserve a single Boolean variable for every qubit and let the satisfying assignments of our formulae represent the state vector. \autoref{ex:variables} illustrates this.
In what follows, we will write $F_{\ket{\phi}}$ for a similar encoding of $\ket{\phi}$. 

\begin{example}
\label{ex:variables}
Let $\ket{B} = \frac1{\sqrt 2}(\ket{01} - \ket{11})$.
Let $x$ be the variable for the first qubit and $y$ be the variable for the second qubit. Written differently, we have:
$\ket{B} =(\frac1{\sqrt 2} \ket{0}_x - \frac1{\sqrt 2} \ket{1}_x)\otimes\ket{1}_y$.
The corresponding Boolean constraint is $F_{\ket{B}} = (x\vee \no x)\wedge y$ where we assign the  $W(x) = -\frac{1}{\sqrt{2}}$ and $W(\no x) = \frac{1}{\sqrt{2}}$, leaving $y$ unbiased.
The (weighted) satisfying assignments are:
$\{
\no xy \equiv \tfrac{1}{\sqrt{2}} \ket{01},
xy \equiv -\tfrac{1}{\sqrt{2}} \ket{11}
\}.$
\hfill$\diamond$
\end{example}

From now on, we will reserve the variables $x,y,z$ for the first three qubits.

\exercise{Encode the so-called Bell state $\tfrac{1}{\sqrt{2}}(\ket{00} + \ket{11})$ in weighted model countning.}

\subsection{Encoding Quantum Gates and Circuits to \#SAT}
\label{sec:encoding-gate}
Since the $NOT$, $CNOT$ and $CCNOT$ (Toffoli) gates are classical (reversible) gates, their encoding is easy. For instance, the Toffoli gate on input bits $x,y,z$ and output bits $x',y',z'$, only flips $z$, i.e., sets $z' \Leftrightarrow \neg z$, when both $x$ and $y$ are set to true, as explained in \autoref{sec:prelims}.
Nothing changes in the quantum setting, except that we reserve the same variables now for the qubits. The Boolean encoding of these gates is thus as follows.
\begin{equation}\label{cons:clifford}
  \begin{aligned}
    &F_{NOT}&(x, x')      \defn   ~&   x' \Leftrightarrow  x \oplus 1 ~~~=~~~ x' \Leftrightarrow  \neg x \\
    &F_{CNOT}&(x,y,x',y') \defn  ~& y' \Leftrightarrow y \oplus x \land x' \Leftrightarrow x\\
    &F_{{CCNOT}}&(x,y,z,x',y',z') \defn ~& 
                z' \Leftrightarrow z \oplus (y \land z) \land x' \Leftrightarrow x \land y'\Leftrightarrow y
  \end{aligned}
\end{equation}
Observe that the above encoding determines the output variables $x',y',z'$ in terms of the input variables, as illustrated in the following example.
\begin{example}
    Let the input state be $\ket{110}$,
    where the corresponding Boolean constraint is $F_{\ket{110}} = x\wedge y\wedge \neg z$.
    After applying $CCNOT$ gate,
    the output state is described by the constraint 
    \[
    F_{\ket{110}}\wedge F_{CCNOT} = (x\wedge y\wedge \neg z) ~~\wedge~~ (z' \Leftrightarrow z \oplus (y \land z) \land x' \Leftrightarrow x \land y'\Leftrightarrow y)
    \]
    with satisfying assignment $x'y'z'$,
    which is interpreted as state $\ket{111}$.
    
    In the (non-classical) case where we have superposition $\frac{1}{\sqrt{2}}(\ket{011}+\ket{111})$,
    each basis state will be a satisfying assignment,
    e.g. $ y z$ with $W(x) = W(\no x) = \frac{1}{\sqrt{2}}$,
    where the satisfying assignments are $\{x y z, \no x y z\}\equiv \frac{1}{\sqrt{2}}(\ket{011}+\ket{111})$. 
    Applying a $CCNOT$ gate ends up with two computational basis states (satisfying assignments) $\{\no x y z, x y\no z\}\equiv \frac{1}{\sqrt{2}}\ket{011} + \ket{110}$.
    \hfill$\diamond$
\end{example}

Recall that its gate semantics in \autoref{ex:h}.
We will encode the gate again as a constraint $F_{H}(x,x',h)$, where 
$x/x'$ is the qubit input/output variable and $h$ is a separate variable representing the $\pm \frac1{\sqrt 2}$ normalization. 
Notice in particular that the encoding should increase the number of satisfying assignments after introducing the $x'$ variable. We achieve this by leaving $x'$ unconstrained.
The following encoding of the Hadamard gate also ensures that the negative weight only occurs for the case when $\ket1$ is the input and $\ket1$ is the output.
\begin{equation}
  \begin{aligned}
    F_{H}(x,x',h) \quad \defn \quad h \Longleftrightarrow  ( x \land x') \quad
\text{   with   } \quad W(\no h) = \frac{1}{\sqrt{2}} \quad W(h) = -\frac{1}{\sqrt{2}}
  \end{aligned}
\end{equation}
\vspace{-2em}
\begin{example}\label{ex:satencoding}
The following circuit (identical to the one used in \autoref{sec:visualizing}) computes the famous Bell state $\ket{\phi_2}=  \frac1{\sqrt 2}(\ket{00} + \ket{11}) $ at time step $2$, only to uncompute it again, ending up back in the $\ket{00}$ state. As will become apparent, this circuit nicely illustrates how the encoding handles constructive and destructive interference.
In the circuit, we explicitly label the qubits with Boolean variables $x,y$ and add a subscript to the Hadamard gates, to indicate that we will reserve an additional Boolean variable $h$ or $h'$ per gate.
 \[
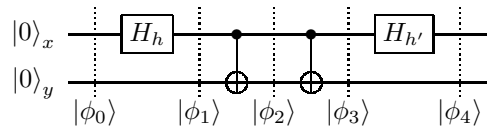
    \begin{array}{c}  
      \Qcircuit @C=1em @R=.7em {
        \lstick{\ket{0}_{ x}} & \qw\ar@{.}[]+<0em,1em>;[d]+<0em,-0.5em> & \gate{H_h} &\qw\ar@{.}[]+<0em,1em>;[d]+<0em,-0.5em> & \ctrl{1} 
        &\qw\ar@{.}[]+<0em,1em>;[d]+<0em,-0.5em> & \ctrl{1} 
        &\qw\ar@{.}[]+<0em,1em>;[d]+<0em,-0.5em> & \gate{H_{h'}}
        &\qw\ar@{.}[]+<0em,1em>;[d]+<0em,-0.5em> & \qw 
        \\
        \lstick{\ket{0}_{ y}} & \qw & \qw & \qw & \targ & \qw   & \targ & \qw & \qw \qw& \qw & \qw
                    \\
        & \ket{\phi_0} & 
        & \ket{\phi_1} &
        & \ket{\phi_2} &
        & \ket{\phi_3} &
        & \ket{\phi_4} &
      }
    \end{array}
\]
We first show the satisfying assignments of the circuit encoding at each time step, where for the encoded Hadamard gates $F_H(x,x',h)$, we add an additional constraint $y \Leftrightarrow y'$ implementing the identity on the second qubit, i.e., $H\otimes I$.
\[
\begin{aligned}
   \ket{\phi_0} \equiv F_{\ket{\phi_0}} &= \neg x_0 \wedge \neg y_0 \hspace{20.5 em}\{\no x_0\no y_0\}\\
    \ket{\phi_1} \equiv F_{\ket{\phi_1}} &= F_{\ket{\phi_0}} \wedge F_{H}( x_0,x_1,h) \land (y_0 \Leftrightarrow y_1)
    \hspace{7em}
    \{\no h x_1 \no y_1, \no h \no x_1 \no y_1 \}
    \\
    \ket{\phi_2} \equiv F_{\ket{\phi_2}} &= F_{\ket{\phi_1}} \wedge F_{CNOT}( x_1,y_1,x_2,y_2)
    \hspace{5em}
    \{\no h x_1 \no y_1 x_2 y_2, \no h\no x_1 \no y_1 \no x_2 \no y_2\}
    \\
    \ket{\phi_3} \equiv F_{\ket{\phi_3}}&= F_{\ket{\phi_2}} \wedge F_{CNOT}( x_2,y_2,x_3,y_3)\quad
    \{\no h x_1 \no y_1 x_2 y_2 x_3 \no y_3, \no h \no x_1 \no y_1 \no x_2 \no y_2 \no x_3 \no y_3\}
\end{aligned}
\]
It is worth noting that,
in the final time step,
the satisfying assignments of
$
    \ket{\phi_4} \equiv  F_{\ket{\phi_4}}  = F_{\ket{\phi_3}}  \wedge F_{H}(x_3,x_4,h') \land (y_3 \Leftrightarrow y_4)
$
will be
\[
\begin{aligned}
    \{ &h' \no h x_1 \no y_1 x_2 y_2 x_3 \no y_3 x_4 \no y_4 \equiv -\tfrac{1}{2}\ket{10},
    &&\no h'\no h  x_1 \no y_1  x_2  y_2 x_3 \no y_3 \no x_4 \no y_4 \equiv \tfrac{1}{2}\ket{00}, \\
    &\no h'\no h \no x_1 \no y_1 \no x_2 \no y_2 \no x_3 \no y_3  x_4 \no y_4 \equiv \tfrac{1}{2}\ket{10}, 
    &&\no h'\no h \no x_1 \no y_1 \no x_2 \no y_2 \no x_3 \no y_3 \no x_4 \no y_4 \equiv \tfrac{1}{2}\ket{00}\}, \\   
\end{aligned}
\]
where we have $(\tfrac{1}{2}-\tfrac{1}{2})\ket{10}$ (destructive interference) and $(\tfrac{1}{2}+\tfrac{1}{2})\ket{00}$ (constructive interference).
Recall in \autoref{sec:visualizing}, we give the transition paths of states in the same circuit.
Consider one of the paths from $\ket{00}$ to itself
corresponding to the satisfying assignment $\no h'\no h \no x_0\no y_0 x_1 \no y_1 x_2 y_2 x_3 \no y_3 \no x_4 \no y_4$:
{
\circledd{$\ket{00}$}{$\no x_0\no y_0$}
$\textcolor{red}{\stackrel[\no h]{\nicefrac1{\sqrt{2}}}{\rightarrow}}$
\circledd{$\ket{10}$}{$x_1\no y_1$}
$\textcolor{blue}{\stackrel{1}{\rightarrow}}$
\circledd{$\ket{11}$}{$x_2 y_2$}
$\textcolor{blue}{\stackrel{1}{\rightarrow}}$
\circledd{$\ket{10}$}{$x_3\no y_3$}
$\textcolor{red}{\stackrel[\no h']{\nicefrac1{\sqrt{2}}}{\rightarrow}}$
\circledd{$\ket{00}$}{$\no x_4\no y_4$}}.
Here we omit the satisfying assignments for $x_0$ and $y_0$ in $SAT(F_{\ket{\psi_t}})$ with $t\in[1,4]$ as each of them contains $\no x_0 \no y_0$.
We see that the satisfying assignments have a one-to-one mapping to the paths.
\hfill$\diamond$
\end{example}

This encoding effectively realizes the well-known path sum approach~\cite{feynman2010quantum} to the classical simulation of quantum circuits. It can be written in linear algebra as follows, where $U_0,U_1\dots, U_m$ are the ($n$-qubit) gates in the circuit and $\ket{\vec b_1}, \ket{\vec b_2}, \dots, \ket{\vec b_m}$ are computational basis states.
\[
   U_m \cdots U_1 U_0 \ket{00\dots0}
   = 
   \sum_{\mathclap{\vec b_1, \vec b_2, \dots, \vec b_m \in \set{0,1}^n}}
    ~U_m~\cdot~ \ketbra{\vec b_m} ~\dots~  \ketbra{\vec b_2}   ~\cdot~U_1~\cdot~ \ketbra{\vec b_1}  ~\cdot~U_0 ~\cdot~ \ket{00\dots0}
\]
The previous example shows that the \#SAT encoding does not ``merge'' paths ending in the same computational basis state, as illustrated below (dashed edges cancel each other out).
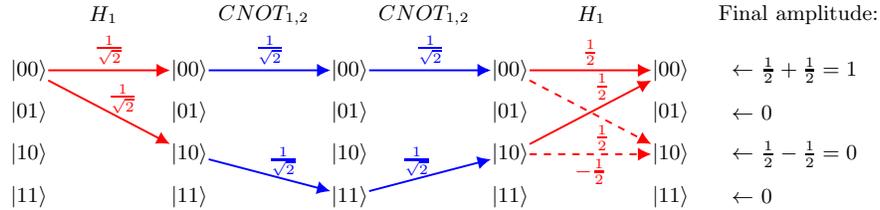
\begin{center}
\scalebox{0.9}{\begin{tikzpicture}[
    scale=0.3,
    every path/.style={>=latex},
    every node/.style={},
    inner sep=0pt,
    minimum size=14pt,
    line width=1pt,
    node distance=1.8cm,
    thick,
    font=\footnotesize
    ]

    \node[] (001)   {$\ket{00}$};
    \node[below =.1cm of 001] (011) {$\ket{01}$};
    \node[below =.1cm of 011] (101) {$\ket{10}$};
    \node[below =.1cm of 101] (111) {$\ket{11}$};
    \node[above = .3cm of 001, xshift = 1.1cm] {$H_1$};
    
    \node[right = of 001] (002) {$\ket{00}$};
    \node[right = of 011] (012) {$\ket{01}$};
    \node[right = of 101] (102) {$\ket{10}$};
    \node[right = of 111] (112) {$\ket{11}$};
    \node[above = .3cm of 002, xshift = 1.1cm] {$CNOT_{1,2}$};

    \node[right = of 002] (003) {$\ket{00}$};
    \node[right = of 012] (013) {$\ket{01}$};
    \node[right = of 102] (103) {$\ket{10}$};
    \node[right = of 112] (113) {$\ket{11}$};
    \node[above = .3cm of 003, xshift = 1.1cm] {$CNOT_{1,2}$};

    \node[right = of 003] (004) {$\ket{00}$};
    \node[right = of 013] (014) {$\ket{01}$};
    \node[right = of 103] (104) {$\ket{10}$};
    \node[right = of 113] (114) {$\ket{11}$};
    \node[above = .3cm of 004, xshift = 1.2cm] (h1) {$H_1$};
        \node[right = 1.6cm of h1] {Final amplitude:};

    \node[right = of 004] (005) {$\ket{00}$};
    \node[right =0.6cm of 005]{$\leftarrow$ $\tfrac{1}{2} + \tfrac{1}{2} = 1$};
    \node[right = of 014] (015) {$\ket{01}$};
     \node[right =0.6cm of 015]{$\leftarrow$ $0$};00
    \node[right = of 104] (105) {$\ket{10}$};
         \node[right =0.6cm of 105]{$\leftarrow$ $\tfrac{1}{2} - \tfrac{1}{2} = 0$};
    \node[right = of 114] (115) {$\ket{11}$};
    \node[right =0.6cm of 115]{$\leftarrow$ $0$};

\draw[red, \bigarrowhead] (001) edge node[yshift=0.3cm] {$\tfrac{1}{\sqrt{2}}$} (002);
\draw[blue, \bigarrowhead] (002) edge node[yshift=0.3cm] {$\tfrac{1}{\sqrt{2}}$} (003);
\draw[blue, \bigarrowhead] (003) edge node[yshift=0.3cm] {$\tfrac{1}{\sqrt{2}}$} (004);
\draw[red, \bigarrowhead] (004) edge node[above,pos=.5] {$\tfrac{1}{2}$} (005);
\draw[red, \bigarrowhead, dashed] (004) edge node[below left=-1mm,pos=.7] {$\tfrac{1}{2}$} (105);
\draw[red, \bigarrowhead, dashed] (104) edge node[below,pos=.5] {$-\tfrac{1}{2}$} (105);

\draw[red, \bigarrowhead] (001) edge node[xshift=0.2cm, yshift=0.2cm] {$\tfrac{1}{\sqrt{2}}$} (102);
\draw[blue, \bigarrowhead] (102) edge node[xshift=0.2cm, yshift=0.2cm] {$\tfrac{1}{\sqrt{2}}$} (113);
\draw[blue, \bigarrowhead] (113) edge node[xshift=-0.2cm, yshift=0.2cm] {$\tfrac{1}{\sqrt{2}}$} (104);
\draw[red, \bigarrowhead] (104) edge node[pos=.7,above left=-1mm] {$\tfrac{1}{2}$} (005);
\end{tikzpicture}
}
\end{center}

\keymessage{
Given a universal quantum circuit, consisting of Toffoli and Hadamard gates, we can construct a Boolean formula whose satisfying assignments represent the circuit's output quantum state.
}

\subsection{Encoding Measurements}
Recall in \autoref{subsec:measurement},
measuring all qubits in computational basis obtains the probability of an outcome $\ket{b}$ for $b\in\{0,1\}^n$.
In a circuit with gates $U_1, \dots, U_m$ and an input state $\ket{\varphi_0}$,
the output state $\ket{\varphi_m} = U_m\cdots U_1\ket{\varphi_0}$ can be decomposed into basis states as $\ket{\varphi_m} = \sum_{b\in\{0,1\}^n}\alpha_{b}\ket{b}$,
where the amplitudes can be computed by weighted model counting
$
\#SAT_W(F_{\ket{\varphi_m}}\wedge F_{\ket{b}}) = \alpha_{b},
$
where $F_{\ket{\varphi_m}} = F_{U_m}\wedge\cdots\wedge F_{U_2} \wedge F_{U_1} \wedge F_{\ket{\varphi_0}}$ follows the encoding in \autoref{sec:encoding-gate} and $F_{\ket{b}}$ is the encoding of basis state $\ket{b}$ as shown in \autoref{sec:encoding-state}.
The probability then equals $(\alpha_{b})^2$.

\begin{example}
Reconsider \autoref{ex:satencoding} with measurements on all qubits of the output state.
To get the probability of measuring basis state $\ket{b} = \ket{00}$,
the constraint
$F_{\ket{b}} = \no x_4\wedge \no y_4$.
should be
conjoined to the final state: $F_{\ket{\varphi_4}}\wedge F_{\ket{b}}$.
We then have $SAT(F_{\ket{\varphi_4}}\wedge F_{\ket{b}}) = \{\no h'\no h  x_1 \no y_1 \no x_2 \no y_2 x_3 \no y_3 \no x_4 \no y_4, \no h'\no h \no x_1 \no y_1 \no x_2 \no y_2 \no x_3 \no y_3 \no x_4 \no y_4\}$.
The amplitude of $\ket{00}$ is $MC_W(F_{\ket{\varphi_4}}\wedge F_{\ket{b}}) = W(h)W(h') + W(h)W(h') = 1$,
thus the resulting probability is $1^2 
= 1$.
\hfill$\diamond$
\end{example}
\exercise{In the above example, compute the probability of measuring with basis state $\ket{01}$.}
By adding the measurement constraint,
only the ``paths'' to the basis state we measure would remain.
In terms of satisfying assignments,
as shown in the previous example,
we will keep the satisfying assignments with variables on the final time step encoding the measured basis state ($\no h'\no h  x_1 \no y_1 \no x_2 \no y_2 x_3 \no y_3 \no x_4 \no y_4\equiv\tfrac{1}{2}\ket{00}, \no h'\no h \no x_1 \no y_1 \no x_2 \no y_2 \no x_3 \no y_3 \no x_4 \no y_4\equiv\tfrac{1}{2}\ket{00}$) and discard others ($h' \no h x_1 \no y_1 x_2 y_2 x_3 \no y_3 x_4 \no y_4 \\\equiv -\tfrac{1}{2}\ket{10}$, 
$\no h'\no h \no x_1 \no y_1 \no x_2 \no y_2 \no x_3 \no y_3  x_4 \no y_4\equiv\tfrac{1}{2}\ket{10}$).

\subsection{Open questions on applications of \#SAT}
We have seen that
 weighted model counting can effectively tackle universal quantum computing.
The above encoding can also be extended beyond Toffoli + $H$ to typical gate sets, like Clifford + $T$ (see \aref{app:gatesets}).
In that case, the variable weights will include complex numbers,
 while modern weighted model counters only support real-valued weights.
 To deal with this problem,
 we can move from computational basis encoding to a different basis, as explained in~\cite{mei2024eq}.
 For interested readers,
 please refer to \aref{app:pauli}.

Future avenues include using approximate model counters for increased speed and exploring satisfiability modulo theories (SMT) tools~\cite{barrett2018satisfiability} for quantum computing.
Initial SMT-based approaches have shown limited scalability~\cite{Bauer2023symQV} because of a reliance on an undecidable theory.
We can also base the \#SAT encoding on density matrices in the Pauli basis~\cite{mei2024simulating} instead of vectors in the computational basis, as above, to obviate the need for complex numbers in the model counting tool.
Finally, exploring whether the weighted model counting approach can also be applied to synthesize quantum circuits with desired properties is worthwhile.

\section{Decision diagrams}
\label{sec:DecisionDiagrams}
\label{sec:dd}

This section introduces decision diagrams and their use in quantum computing. We will see that decision diagrams can be regarded as a compression technique for quantum states (vectors). Identical parts of the vector are reused multiple times. This leads to an exponential space reduction of the vector in the best case. 
Some figures and examples in this section were adapted from~\cite{Thanos2024automated,limdd}.
The main concepts were introduced earlier~\cite{akers1978binary,bryant86,miller2006qmdd}.

\subsection{What are decision diagrams?}
In classical computing, decision diagrams are used to represent Boolean functions compactly, for example, in Boolean decision diagrams (BDD). They have been successfully applied to satisfiability, reachability and model checking since their introduction in 1978~\cite{akers1978binary,bryant86}.
We will show that decision diagrams can also be used for the quantum domain.

As shown in \autoref{sec:quantumComputingFromCSperspective}, vectors can represent quantum states. The size of such vectors grows exponentially in the number of qubits. Hence, a quantum state with only 60 qubits requires a few exabytes ($=10^3\text{ petabytes } = 10^6\text{ terabytes}$) of space. 
We explain how a decision diagram often compactly represents a quantum state, while allowing for efficient operations on them. Assume we have some quantum state $\ket{\psi}$ on $n$ qubits, which we can equally represent as a pseudo-Boolean function $f:\{0,1\}^n\to \com$ as follows.

\begin{equation}
    \ket{\psi} = \sum_{\vec{x}\in\{0,1\}^n}f(\vec{x})\ket{\vec{x}} =
    \begin{bmatrix}
        f(00\dots00)\\
        f(00\dots01)\\
        \vdots\\
        f(11\dots10)\\
        f(11\dots11)
    \end{bmatrix}
\end{equation}

Now we decompose the state $\ket{\psi}$ in the top half of the vector and the bottom half of the vector. In terms of pseudo-Boolean functions, we use the Shannon decomposition $f = (x\wedge f_x)\vee (\overline{x}\wedge f_{\overline{x}})$, where $f_x, f_{\overline{x}}$ are the functions with variable $x$ set to 1, 0. In this paper, we will change the notation to $f_0:=f_{\overline{x}}$ and $f_1:=f_x$ where the variable $x$ is always the first variable of $\vec{x}$. In other words, we have $f_0(\vec{x}) = f(0, \vec{x})$ and $f_1(\vec{x}) = f(1, \vec{x})$.
We define $\ket{\psi_0}$ as $\ket{\psi_0} = \sum_{\Vec{x}\in\{0,1\}^{n-1}}f_0(\Vec{x})\ket{\vec{x}}$ and similarly $\ket{\psi_1} =  \sum_{\Vec{x}\in\{0,1\}^{n-1}}f_1(\Vec{x})\ket{\vec{x}}$.
The possibly unnormalized states $\ket{\psi_0}$ and $\ket{\psi_1}$ will be called substates of $\ket{\psi}$. In terms of vectors, we now have the following decomposition (left):

\begin{center}
\begin{tikzpicture}
    \node[](equation){
        $
            \ket{\psi} = \begin{bmatrix}
                f(00\dots00)\\
                \vdots\\
                f(01\dots11)\\
                f(10\dots00)\\
                \vdots\\
                f(11\dots11)
            \end{bmatrix}
            = 
            \begin{bmatrix}
                \psi_0\\
                \psi_1
            \end{bmatrix}
            = \ket{0}\otimes\ket{\psi_0} + \ket{1}\otimes\ket{\psi_1}
        $
    };

    \node[right= 0cm of equation](diagram){
        \begin{tikzpicture}[
            scale=0.3,
            every path/.style={>=latex},
            every node/.style={},
            inner sep=1pt,
            minimum size=0.3cm,
            line width=1pt,
            node distance=.8cm,
            thick,
            font=\scriptsize
            ]
                \node[circle,draw]      (top)                              {$\ket{\psi}$};
                \node[circle] (down) [below=of top] {};
                \node[circle,draw]        (left)       [left=of down] {$\ket{\psi_0}$};
                \node[circle,draw]      (right)       [right=of down] {$\ket{\psi_1}$};
                \node[circle]        (toptop)       [above=of top] {};
                
                \draw[->,e1=0] (toptop) -- (top);
                \draw[->,e0=0] (top) -- (left);
                \draw[->,e1=0] (top) -- (right);
                
        \end{tikzpicture}
    
    };

\end{tikzpicture}
\end{center}

In terms of decision diagrams, this state is represented by the diagram right. (Note the one-to-one correspondence between a decision diagram node, a substate, a vector and a pseudo-Boolean function.)
A dotted edge is a so-called zero-edge (pointing to $\ket{\psi_0}$) and a solid edge is a one-edge (pointing to $\ket{\psi_1}$).

We can recurse this procedure for every new substate. At the base case of the recursion at the low level, the node consists of a complex valued coefficient, representing corresponding vector entry (see the figure below).
Thus, in every layer of the decision diagram, one variable is consumed and split into True and False cases. This leads to a decision tree.

Note that this decision tree is exponentially large: a decision tree of depth $n$ has $2^{n}-1$ interior nodes and $2^{n-1}$ leaves. But often the size of this decision tree can be reduced using dynamic programming: 
if two nodes represent the same function,
they can be merged. Stated differently, two nodes that represent the same quantum substate can be merged.
This leads to a decision diagram.

\begin{wrapfigure}{r}{.6\textwidth}
\vspace{-.8cm}
\tikzset{every picture/.style={->,thick}}
\centering
\begin{tikzpicture}[
    scale=0.3,
    every path/.style={>=latex},
    every node/.style={},
    inner sep=1pt,
    minimum size=0.3cm,
    line width=1pt,
    node distance=.8cm,
    thick,
    font=\scriptsize
    ]
    \colorlet{green}{ForestGreen}
    \colorlet{blue}{RoyalBlue}
    \colorlet{purple}{Orchid}
    \colorlet{red}{Red}

    \node[draw,circle] (a1) {$x$};
    \node[draw,circle, below = .4cm of a1, xshift=-.65cm] (a2) {$y$};
    \node[draw,circle, below = .4cm of a1, xshift= .65cm] (a3) {$y$};
    
    \node[draw,circle, below = .4cm of a2, xshift=-.3cm,fill={red}] (a41) {$z$};
    \draw[e0=  0] (a2) edge  node[] {} (a41);
    \node[draw,circle, below = .4cm of a2, xshift= .3cm,fill=purple] (a42) {$z$};
    \node[draw,circle, below = .4cm of a3, xshift=-.3cm,fill=green] (a43) {$z$};
    \node[draw,circle, below = .4cm of a3, xshift= .3cm,fill=blue] (a44) {$z$};

    \node[leaf, below=.35cm of a41, xshift=-0cm      ] (w1) {$1$};
    \node[leaf, below=.35cm of a42,inner sep=0pt] (w3) {$2$};
    \node[leaf, below=.35cm of a43,inner sep=0pt] (w5) {$1$};
    \node[leaf, below=.35cm of a44,inner sep=0pt] (w7) {$-2$};

    \draw[<-] (a1) --++(90:2cm) node[right,pos=.7] {$f$};
    \draw[e0 = 0] (a1) edge  node[] {} (a2);
    \draw[e1 = 0] (a1) edge  node[] {} (a3);

    \draw[e1=  0] (a2) edge  node[] {} (a42);
    \draw[e0=  0] (a3) edge  node[] {} (a43);
    \draw[e1=  0] (a3) edge  node[] {} (a44);

    \draw[e0=  0] (a41) edge  node[] {} (w1);
    \draw[e0=  0] (a42) edge  node[] {} (w3);
    \draw[e0=  0] (a43) edge  node[] {} (w5);
    \draw[e0=  0] (a44) edge  node[] {} (w7);

    \node[below= 2.3cm of a1]   (dt)  {Decision tree};

\node[left = 1.4cm of a1.south,yshift=-1cm] (vec) {
    \begin{minipage}{1.5cm}\scriptsize
    $\def\arraystretch{1.3}
    \begin{matrix*}[l]
	{000:}\\
        {001:}\\
        {010:}\\
        {011:}\\
        {100:}\\
        {101:}\\ 
        {110:}\\
        {111:}\\
    \end{matrix*}
    \begin{bmatrix*}[r]
	{\color{red} 1}\\
        {\color{red} 0}\\
        {\color{purple} 2}\\
        {\color{purple} 0}\\
        {\color{green} 1}\\
        {\color{green} 0}\\ 
        {\color{blue}-2}\\
        {\color{blue} 0}\\
    \end{bmatrix*}$
    \end{minipage}
};

    \node[draw,circle, right=2.5cm  of a1] (a1) {$x$};
    \node[draw,circle, below = .4cm of a1, xshift=-.65cm] (a2) {$y$};
    \node[draw,circle, below = .4cm of a1, xshift= .65cm] (a3) {$y$};
    
    \node[draw,circle, below = .4cm of a2, xshift= 0cm,fill=purple] (a42) {$z$};
    \node[draw,circle, below = .4cm of a3, xshift=-.3cm, shade, shading=axis, left color=red,  middle color=red, right color=green, shading angle=90] (a43) {$z$};
    \node[draw,circle, below = .4cm of a3, xshift= .3cm,fill=blue] (a44) {$z$};

    \node[leaf, below=.35cm of a42,inner sep=0pt] (w3) {$2$};
    \node[leaf, below=.35cm of a43,inner sep=0pt] (w5) {$1$};
    \node[leaf, below=.35cm of a44,inner sep=0pt] (w7) {$-2$};

    \draw[<-] (a1) --++(90:2cm) node[right,pos=.7] {$f$};
    \draw[e0 = 0] (a1) edge  node[] {} (a2);
    \draw[e1 = 0] (a1) edge  node[] {} (a3);

    \draw[e0=  0] (a2) edge  node[] {} (a43);
    \draw[e1=  0] (a2) edge  node[] {} (a42);
    \draw[e0=  0] (a3) edge  node[] {} (a43);
    \draw[e1=  0] (a3) edge  node[] {} (a44);
    \draw[e0=  0] (a42) edge  node[] {} (w3);
    \draw[e0=  0] (a43) edge  node[] {} (w5);
    \draw[e0=  0] (a44) edge  node[] {} (w7);

    \node[below= 2.3cm of a1]   (dt)  {ADD};

\end{tikzpicture}
\vspace{-0.5cm}
\end{wrapfigure}

The figure on the right (taken from~\cite{Thanos2024automated}) shows an example of node merging in decision diagrams. It displays a decision tree (middle) and decision diagram (right) for the same function $f:\{0,1\}^3\to\mathbb{C}$, depicted as a vector on the left. As the red and green nodes in the decision tree represent the same function (or substate) $\begin{bsmallmatrix}
    1\\
    0
\end{bsmallmatrix}$, they are merged in the decision diagram. The decision diagram that merges nodes that represent the same function is called the Algebraic Decision Diagram (ADD) (other decision diagram types are discussed in the next subsection).

The following exercise shows that node merging can exponentially reduce decision diagrams. 
\exercise{Check that the quantum states $\frac{1}{2}[1,1,1,1,0,0,0,0]$ and $\ket{\psi}=\sum_{{b}\in\{0,1\}^n}\frac{1}{\sqrt{2^n}}\ket{{b}}$ have a decision tree with $2^n -1$ interior nodes, but their ADDs have $n$ interior nodes.}

\keymessage{The type of decision diagram as explained above forms a compression technique for quantum states that compress states based on the equality of their substates.
}

\subsection{Types of decision diagrams}

As explained above, nodes in a decision diagram are merged if they represent the same states. There exist more general methods for merging nodes leading to other types of decision diagrams. For example, in the Quantum Multiple-Valued Decision Diagram (QMDD) two states are merged if they are equal up to a constant. So states $\ket{\psi_0}$ and $\ket{\psi_1}$ are merged if their corresponding pseudo-Boolean functions obey $f_0=c\cdot f_1$ for some constant complex number $c$. The constant $c$ is placed on the edge in the decision diagram such that the original state can be identified. QMDD gives more compression of the state vector with respect to ADD. This compression is exponential in the best case as this exercise shows: \exercise{Show that the vector $[1,2,2,4,4,8,8,16]$ and the function $f(x_1,\dots,x_n)=\prod_{k=1}^n (p_k)^{x_k}$ (where $p_k$ is the $k$-th prime number) have $2^n-1$ ADD nodes, but their QMDDs have only $n+1$ nodes.}

\begin{wrapfigure}{R}{.55\textwidth}
\vspace{-.8cm}
\tikzset{every picture/.style={->,thick}}
\centering
\begin{tikzpicture}[
    scale=0.3,
    every path/.style={>=latex},
    every node/.style={},
    inner sep=1pt,
    minimum size=0.3cm,
    line width=1pt,
    node distance=.8cm,
    thick,
    font=\scriptsize
    ]

\colorlet{origgreen}{green}
\colorlet{origblue}{blue}
\colorlet{origpurple}{purple}
\colorlet{origred}{red}

\colorlet{green}{ForestGreen}
\colorlet{blue}{RoyalBlue}
\colorlet{purple}{Orchid}
\colorlet{red}{Red}

\colorlet{origgreen}{green}
\colorlet{origblue}{blue}
\colorlet{origpurple}{purple}
\colorlet{origred}{red}

\colorlet{green}{ForestGreen}
\colorlet{blue}{RoyalBlue}
\colorlet{purple}{Orchid}
\colorlet{red}{Red}

    \node[circle] (a1) {};

\node[left = 1.4cm of a1.south,yshift=-1cm] (vec) {
    \begin{minipage}{1.5cm}\scriptsize
    $\def\arraystretch{1.3}
    \begin{matrix*}[l]
	{000:}\\
        {001:}\\
        {010:}\\
        {011:}\\
        {100:}\\
        {101:}\\ 
        {110:}\\
        {111:}\\
    \end{matrix*}
    \begin{bmatrix*}[r]
	  {\color{green} 1}\\
        {\color{green} 0}\\
        {\color{purple} 2}\\
        {\color{purple} 0}\\
        {\color{green} 1}\\
        {\color{green} 0}\\ 
        {\color{blue}-2}\\
        {\color{blue} 0}\\
    \end{bmatrix*}$
    \end{minipage}
};

    \node[draw,circle, right = -.5cm  of a1] (a1) {$x$};
    \node[draw,circle, below = .4cm of a1, xshift=-.65cm] (a2) {$y$};
    \node[draw,circle, below = .4cm of a1, xshift= .65cm] (a3) {$y$};
    
    \node[draw,circle, below = .4cm of a2, xshift= 0cm,fill=purple] (a42) {$z$};
    \node[draw,circle, below = .4cm of a3, xshift=-.3cm,fill=green] (a43) {$z$};
    \node[draw,circle, below = .4cm of a3, xshift= .3cm,fill=blue] (a44) {$z$};

    \node[leaf, below=.35cm of a42,inner sep=0pt] (w3) {$2$};
    \node[leaf, below=.35cm of a43,inner sep=0pt] (w5) {$1$};
    \node[leaf, below=.35cm of a44,inner sep=0pt] (w7) {$-2$};

    \draw[<-] (a1) --++(90:2cm) node[right,pos=.7] {$f$};
    \draw[e0 = 0] (a1) edge  node[] {} (a2);
    \draw[e1 = 0] (a1) edge  node[] {} (a3);

    \draw[e0=  0] (a2) edge  node[] {} (a43);
    \draw[e1=  0] (a2) edge  node[] {} (a42);
    \draw[e0=  0] (a3) edge  node[] {} (a43);
    \draw[e1=  0] (a3) edge  node[] {} (a44);
    \draw[e0=  0] (a42) edge  node[] {} (w3);
    \draw[e0=  0] (a43) edge  node[] {} (w5);
    \draw[e0=  0] (a44) edge  node[] {} (w7);

    \node[below= 2.4cm of a1]   (dt)  {ADD};

    \node[draw,circle, right=2.3cm  of a1] (a1) {$x$};
    \node[draw,circle, below = .4cm of a1, xshift=-.85cm] (a2) {$y$};
    \node[draw,circle, below = .4cm of a1, xshift= .85cm] (a3) {$y$};

    \pgfdeclarehorizontalshading{grad}{100bp}{
        color(0bp)=(purple);
        color(5bp)=(purple);
        color(25bp)=(green);
        color(45bp)=(blue);
        color(50bp)=(blue)
    }

    \node[draw,circle, below = .4cm of a2, xshift=.85cm,shading=grad,shading angle=0] (a43) {$z$};

    \node[leaf, below=.35cm of a43      ] (w5) {$1$};

    \draw[<-] (a1) --++(90:2cm) node[right,pos=.7] {$f$};
    \draw[e0 = 0] (a1) edge  node[] {} (a2);
    \draw[e1 = 0] (a1) edge  node[] {} (a3);

    \draw[e0=15] (a2) edge  node[below left] {$1$} (a43);
    \draw[e1=15] (a2) edge  node[above right] {$2$} (a43);
    \draw[e0=20] (a3) edge  node[above left] {$1$} (a43);
    \draw[e1=20] (a3) edge  node[right] {$-2$} (a43);
    \draw[e0=  0] (a43) edge  node[] {} (w5);

    \node[below= 2.4cm of a1]   (dt)  {QMDD};

\end{tikzpicture}
\vspace{-0.8cm}
\end{wrapfigure}

The diagrams on the right show how states are merged in QMDD. The figure displays an ADD (middle) and QMDD (right) for the same function $f:\{0,1\}^3\to\com$, depicted as a vector on the left. The green node represents the function $f_{green}(x_3)=\overline{x_3}$, the purple node represents $f_{purple}(x_3)=2\cdot\overline{x_3}$ and the blue node $f_{blue}(x_3)=-2\cdot\overline{x_3}$. Thus, all functions in the last layer are equal up to a constant.
Hence, these nodes are merged, and the constants 1, 2, -2 are placed on the edges.

The space gain of decision diagrams is solely due to the ability to merge nodes that represent equivalent functions (for ADD/QMDD, equivalent means equal/equal up to a constant). To efficiently merge nodes, a unique representation of equivalent nodes is needed. By adding constraints to the decision diagram, we ensure that a single node represents each equivalence class of functions. 
For example, for QMDDs, a rule to make each node canonical is to enforce that the low edge label is 1 (this rule is, for example, applied in the QMDD of the figure above). Doing so ensures that the high-edge label is canonical too.
Due to this unique representative, the time for merging nodes decreases from quadratic to linear in the number of diagram nodes: one no longer needs to check equivalence for each pair of nodes but can instead make a node canonical and subsequently check, for each given node, whether it already exists in the diagram.

\begin{table}[t!]
\caption{
Various decision diagrams (DDs) used in the literature (reproduced from~\cite{Thanos2024automated}).
The column ``node merge'' lists the conditions under which two decision diagram nodes, representing functions $f_0$ and $f_1$, are merged. 
Here, $p,a\in \mathbb C$ are complex constants, $P = P_1\otimes \dots \otimes P_n$ a sequence of single-qubit gates $P_i$, and
$f + a$ means the function $f(\vec x) + a$ for all~$\vec x$.
All DDs
represent functions $f_0,f_1\colon \{0,1\}^n\to\mathbb C$.
}
\label{tab:dds}
\centering\scriptsize
\def\arraystretch{1.3}
\begin{tabular}{|p{10cm}|c|}
\hline
\textbf{(Quantum) decision diagrams (and variants)}				& \textbf{Node merge}   \\\hline
Decision Tree	& (no merging)  \\
MTBDD (1993)~\cite{clarke1993spectral}, ADD (1997)~\cite{bahar1997algebric}, QuiDD (2003)~\cite{viamontes2003improving}			& $f_0 = f_1$   \\
SLDD$_\times$\hspace{-.5mm} (2004)~\cite{wilson2005decision}, QMDD (2006)~\cite{miller2006qmdd}, XQDD (2008)~\cite{wang2008xqdd}, TDD (2021)~\cite{hong2022tensor}
	& $f_0 = p\cdot f_1$   \\
EVBDD (1994)~\cite{lai1994evbdd}, SLDD$_+$ (2005)~\cite{fargier2013semiring} & $f_0 = f_1 + a$   \\
FEVBDD (1994)~\cite{tafertshofer1994factored,tafertshofer1997factored,Vrudhula1996},
    SLDD$_{+,\times}$ (2005)~\cite{fargier2013semiring}, AADD (2005)~\cite{aadd} 		& $f_0 = p\cdot f_1 + a$  \\
LIMDD (2023)~\cite{limdd,vinkhuijzen2023efficient,vinkhuijzen2024a}			& $f_0 = p  \cdot  P  \cdot f_1$    \\
\hline
\end{tabular}
\end{table}

Other types of node merging also exist, leading to other types of decision diagrams. \autoref{tab:dds} contains an overview of decision diagrams with their corresponding node merge strategies. We now briefly discuss their distinctive properties.

A type of decision diagram that requires special attention is Local Invertible Map Decision Diagram (LIMDD) as it has provable exponential separation with respect to QMDD: LIMDDs can efficiently represent all states reachable by Clifford circuits~\cite{aaronson2008improved}, which are central in quantum computing, while ADD and QMDD require exponential space for some of those. 
In LIMDD two states are merged if the vectors they represent are equal up to multiplication by a tensor product of single-qubit gates (i.e. `local' gates).
Merging two nodes and calculating the unique edge labels become a polynomial-time calculation (while for ADD and QMDD, this is constant time; see the previous paragraph on canonicity).\footnote{One can check whether the canonical node already exists using a constant-time hash function.} 
Thus calculating the canonical state representation is slower for LIMDD than for ADD and QMDD compared to the size of the decision diagram. 
But as a LIMDD representation of a state is exponentially smaller than for ADD and QMDD in best case, using LIMDD can thus still be more efficient.

\keymessage{There is a zoo of different decision diagram types, each with its strengths and weaknesses in terms of succinctness and tractability of operations. 
}

\subsection{Simulating quantum circuits with decision diagrams}

Like vectors, quantum gates (as matrices) can be represented by decision diagrams. In a decision diagram, a matrix is decomposed in the first decision diagram layer ($x$) in an upper and lower half and in the next layer ($x'$) in a left and right half. Thus a matrix is decomposed into its four quadrants. This decomposition is recursed for every submatrix.

\begin{example}\label{ex:H-gate-as-DD}
    (Left) We can write a $2^n \times 2^n$ matrix $U$ as ADD by decomposing it into four quadrants: the $2^{n-1} \times 2^{n-1}$ matrices $u_{00}, u_{01}, u_{10}$ and $u_{11}$.
    (Right) ADD for the $H$ gate.
    
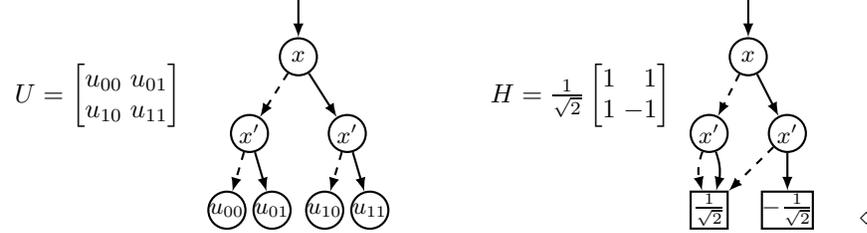
\begin{center}
    \vspace{-1cm}
    \begin{tikzpicture}
        \node[] (vector) {$H = \frac{1}{\sqrt{2}}\begin{bmatrix*}[r]
            1&1\\
            1&-1\\
        \end{bmatrix*}$};

        \node[] (DD) [right=of vector,xshift=-1cm] {
        \begin{tikzpicture}[
            scale=0.3,
            every path/.style={>=latex},
            every node/.style={},
            inner sep=0pt,
            node distance=.5cm,
            minimum size=14pt,
            line width=1pt,
            thick,
            font=\footnotesize
            ]
            \node[circle,draw](top){$x$};
            \node[](toptop)[above=of top]{};
            
            \node[circle](l1)[below=of top]{};
            \node[circle,draw](left1) [left=of l1,xshift=0.5cm]{$x'$};
            \node[circle,draw](right1) [right=of l1,xshift=-0.5cm]{$x'$};
            
            \node[circle](l2) [below=of l1]{};
            \node[rectangle,draw](left2) [below=of left1]{$\frac{1}{\sqrt{2}}$};
            \node[rectangle,draw](right2) [below=of right1]{$-\frac{1}{\sqrt{2}}$};

            \draw[->,e1=0](toptop)--(top){};
            
            \draw[->,e0=0] (top) to (left1){};
            \draw[->,e1=0] (top) to (right1){};

            \draw[->,e0=20] (left1) to  (left2){};
            \draw[->,e1=20] (left1) to  (left2){};

            \draw[->,e0=0](right1)--(left2){};
            \draw[->,e1=0](right1)--(right2){};
            
        \end{tikzpicture}
        };

        \node[left=of vector](DD2){
        \begin{tikzpicture}[
            scale=0.3,
            every path/.style={>=latex},
            every node/.style={},
            inner sep=0pt,
            node distance=.5cm,
            minimum size=14pt,
            line width=1pt,
            thick,
            font=\footnotesize
            ]
            \node[draw,circle] (r) {$x$};
            \node[draw,circle,below = of r,xshift=-.65cm] (a1) {$x'$};
            \node[draw,circle,below = of r,xshift= .65cm] (a2) {$x'$};
            \node[above=of r](toptop){};
          
            \node[draw,circle,below = of a1,xshift=-.3cm] (a11) {$u_{00}$};
            \node[draw,circle,below = of a1,xshift= .3cm] (a12) {$u_{01}$};
            \node[draw,circle,below = of a2,xshift=-.3cm] (a21) {$u_{10}$};
            \node[draw,circle,below = of a2,xshift= .3cm] (a22) {$u_{11}$};
        
            \draw[->,e1=0] (toptop) edge node[left] {} (r);
            
            \draw[->,e0=0] (r) edge  node[left]  {} (a1);
            \draw[->,e1=0] (r) edge  node[right] {} (a2);
            \draw[->,e0=0] (a1) edge  node[left]  {} (a11);
            \draw[->,e1=0] (a1) edge  node[right] {} (a12);
            \draw[->,e0=0] (a2) edge  node[left]  {} (a21);
            \draw[->,e1=0] (a2) edge  node[right] {} (a22);  
        \end{tikzpicture}
        };

        \node[left=of DD2, xshift=1cm] (uMatrix) {$U =         \begin{bmatrix}
            u_{00} &  u_{01}   \\
                u_{10} & u_{11} \\
        \end{bmatrix}$
        };
    \end{tikzpicture}
    \end{center}
    \vspace{-1cm}
    \hfill$\diamond$
\end{example}

\vspace{-.2cm}
In quantum simulation, we want to multiply a matrix (gate) with a vector (state). Note that in decision diagrams a state $\ket{\psi}$ and gate $U=\begin{bsmallmatrix}
    u_{00} & u_{01}\\
    u_{10} & u_{01}
\end{bsmallmatrix}$ 
where each $u_{i,j}$ is again a matrix, can be represented as follows.
\begin{align}
\label{eq:u-dd}
U &= \ket{0}\bra{0}\otimes u_{00} + \ket{0}\bra{1}\otimes u_{01} + \ket{1}\bra{0}\otimes u_{10} + \ket{1}\bra{1}\otimes u_{11}\\
\label{eq:psi-dd}
\ket{\psi} &= \ket{0}\otimes\ket{\psi_0} + \ket{1}\otimes\ket{\psi_1}.
\end{align}
The updated state $U\ket{\psi}$ can thus be written as
\begin{equation}
\label{eq:u-psi}
    U\ket{\psi} = 
    \ket{0}\otimes(u_{00}\ket{\psi_{0}} + u_{01}\ket{\psi_{1}})+
    \ket{1}\otimes(u_{10}\ket{\psi_{0}} + u_{11}\ket{\psi_{1}})
\end{equation}
\exercise{Derive \autoref{eq:u-psi} from \autoref{eq:u-dd} and \autoref{eq:psi-dd} using the fact that
$\bra{a}\cdot\ket{b}$ for $a,b\in\{0,1\}$ equals 1 if $a=b$ and 0 if $a\not=b$. 
}
This formula leads to a recursive algorithm. Note that adding two vectors is also recursive in decision diagrams as can be observed from the following formula:
$$\ket{\psi}+\ket{\varphi} = \ket{0}\otimes(\ket{\psi_0}+\ket{\varphi_0})+\ket{1}\otimes(\ket{\psi_1} + \ket{\varphi_1}).$$
Thus, using dynamic programming, the matrix-vector product is calculated.

\new{To simulate measurement on the qubits, we want to know the probability of an outcome $\ket{b}$ for $b\in\{0,1\}^n$. As we saw in~\autoref{subsec:measurement}, the outcome is $|\alpha_b|^2$. To find this value, we can follow the path representing $b$ and multiply all the values on the edges (for QMDD and LIMDD) or look-up the value in the terminal node of the path (for ADD). Hence the measurement probabilities can be efficiently calculated.}

The multiplication and measurement techniques described above hold for all types of decision diagrams. The difference in performance between the decision diagram types is caused by the efficiency of simulating a specific circuit. As we saw before, Clifford circuits can produce states that are exponentially as large as QMDD but as small as LIMDD. Consequently, LIMDD can efficiently simulate every Clifford circuit, while QMDDs can become exponentially sized when simulating these circuits. Moreover, circuits built from the universal gate set Clifford+$T$ can be simulated by LIMDD in space and time exponentially bounded by the number of $T$-gates. These considerations can be summarized in a so-called quantum knowledge compilation map, which shows for which decision diagrams there are exponential separations between representations of the same state or circuit~\cite{vinkhuijzen2024a}.

Another application of decision diagrams is quantum circuit equivalence checking. By multiplying its gates, a circuit can be represented as a decision diagram. Similarly, we can also represent the inverse $V^{-1}$ of a circuit $V$. The product of a circuit $U$ with the inverse of another circuit $V$ equals the identity $UV^{-1}=I^{\otimes n}$ (possibly up to a complex constant) if and only if the circuits $U$ and $V$ are equivalent. Thus equivalence checking of two circuits $U,V$ consists of applying the multiplication $UV^{-1}$ in the decision diagram~\cite{burgholzer2020advanced}.

Other applications of decision diagrams in the quantum domain are quantum state preparation~\cite{mozafari2022efficient}, quantum circuit optimization, quantum circuit synthesis~\cite{zulehner2017improving} and approximate versions of the above tasks. The methods using decision diagrams turn out to be competitive with other state-of-the-art methods.

\keymessage{
Decision diagrams are currently used in the quantum domain for quantum circuit simulation, quantum circuit equivalence checking and other tasks. They are competitive with state-of-the-art methods for these tasks.
}

\subsection{Open questions on applications of decision diagrams}

In practice, finite floating point precision might lead to two nodes representing an equivalent function no longer being recognized as equivalent. As a consequence, two nodes that should be merged in theory are not done so in real-world implementations,
leading to a blow-up of the decision diagram size or (when a merging error is allowed) a loss of accuracy.
Despite investigations into this trade-off~\cite{niemann2020overcoming}, there currently exists no general solution.

Another direction for further research is the design of new decision diagram types, such as the recent CFLOBDD, where nodes may share substates on different variables~\cite{sistla2023weighted}.

\section{Wrap-up}
\label{sec:outro}

This tutorial detailed how the basic task of \emph{classical simulation} of quantum computing can be done using \#SAT or decision diagrams, which illustrates the inherent combinatorial nature of quantum computing as well as interference, which amplifies or cancels amplitudes of the output quantum state.
We now explain some important open problems in quantum computing and demonstrate, based on related work, how similar methods can solve them.

Current quantum computers will have limited numbers of qubits that are noisy~\cite{preskill2018quantum}.
For this reason, investment in the field only took off after the invention of quantum error correction~\cite{gottesman1997stabilizer}, which realizes the ideal quantum circuit model on noisy hardware.
However, the added complexity of error correction requires additional resources since the logical error-corrected qubits consist of many physical qubits.
So to attain a quantum advantage earlier, we need to optimize quantum algorithms to use few qubits and respect the gate sets and topology (physical connectivity of qubits) of the many different quantum hardware types. 

Therefore, we need to optimize quantum circuits,
synthesize them in different gate sets and verify their correctness. The latter starts with checking whether an optimized circuit implements the same operations as its original (equivalence checking), but also involves checking quantum Hoare logic~\cite{ying2012floyd}.
Finally, error-corrected quantum computers will work in tandem with classical computers to monitor the process~\cite{fowler2012surface}, which requires solving tasks strongly related to the ones discussed above.

Formal methods can indeed be extended to solve the above tasks.
For instance, decision diagrams have been used to check the equivalence of  quantum circuits~\cite{burgholzer2020advanced,viamontes2007equivalence} 
 and model counting~\cite{mei2024eq} as well.
 For a more detailed overview of the successes of formal methods in this domain, 
we refer readers to a historical overview~\cite{Thanos2024automated}
(in which all authors of this tutorial were involved).
As an honorable mention, we note that there are also approaches based on diagrammatic reasoning using the ZX calculus~\cite{coecke2011interacting,wetering2020zx}, which has also been used for circuit optimization~\cite{kissinger2020reducing}, on planning~\cite{shaik2023optimal} and on abstract interpretation~\cite{Bichsel_2023}.

However, the applications of our methods can be viewed in an even broader context. The features that make quantum compilation difficult are shared with the computationally-hard aspects of many problems in quantum physics and quantum chemistry, such as computing the energy of the ground state of a physical system~\cite{kempe2006complexity} or simulating a many-body system~\cite{ORUS14}.
Indeed, many of these problems in physics are in the quantum analogs of the complexity classes \NP{} and~\P, i.e., in \QMA~\cite{bookatz2012qma} and \BQP~\cite{kitaev2002classical}; see the figure below for how these quantum complexity classes relate to classical ones (reproduced from~\cite{deshpande2022importance}).
Therefore, any progress in tackling hard problems for quantum circuits can be directly used to solve the very problems that quantum physicists and chemists struggle with on a daily basis (using the default reductions between \BQP- and \QMA-complete problems).
This is important since even if the ideal error-corrected quantum computer never gets built, we are still stuck with the myriad of ``quantum-hard'' problems that nature poses --- the very reason why Feynman proposed building a quantum computer by inverting the problem in the first place (see \autoref{sec:intro}).

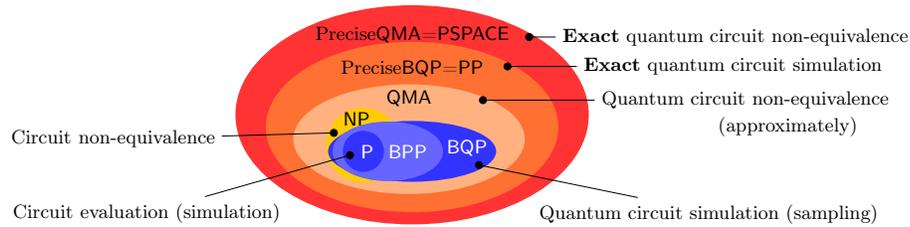
\scalebox{.9}{
\scalebox{0.9}{\tikzset{venn circle/.style={circle,minimum width=2cm,fill=#1,opacity=0.4}}
\begin{tikzpicture}[
e/.style = {ellipse, draw=none,
            minimum height=.7cm, minimum width=.9cm, fill=#1, inner sep=0pt},
e/.default=black
]
\footnotesize

  \node [e = yellow!0!red!80!white,text width=2.8cm,align=center, minimum width=5.8cm,  minimum height=3.6cm,text opacity=1,label={[anchor=north, inner sep=3pt,xshift=-0.cm,yshift=-.25cm]north:Precise\QMA{}=\PSPACE}] (PSPACE) at (.8,0.5) {};
	\node[right = of PSPACE.north east,xshift=-.7cm] (ECEQ) {\color{black} \textbf{Exact} quantum circuit non-equivalence};
	\draw[-Circle,shorten >= -.2cm] (ECEQ) -- (PSPACE.north east);

  \node [e = yellow!30!red!80!white,text width=2.3cm,align=center, minimum width=4.8cm,  minimum height=2.8cm,text opacity=1,label={[anchor=north, inner sep=3pt,xshift=-0.cm, yshift=-.2cm]north:Precise\BQP{}=\PP}] (PP) at (.8,0.3) {};
	\node[right = of PP.north east] (EQCSIM) {\color{black} \textbf{Exact} quantum circuit simulation};
	\draw[-Circle,shorten >= -.2cm] (EQCSIM) -- (PP.north east);

  \node [e = yellow!40!red!50!white,text width=2cm,align=center, minimum width=3.8cm,  minimum height=1.8cm,text opacity=1,label={[anchor=north, inner sep=3pt]north:\QMA}] (QMA) at (.75,0.1) {};
	\node[right = of QMA.north east,xshift=.7cm] (CEQ) {\color{black} Quantum circuit non-equivalence};
	\node[below= of CEQ,yshift=1.1cm,xshift=.7cm] {\color{black} (approximately)};
	\draw[-Circle,shorten >= -.2cm] (CEQ) -- (QMA.north east);

	\node [venn circle = yellow!80!red,text width=1cm,align=center, minimum width=.8cm,opacity=.5,text opacity=1,label={[anchor=west, inner sep=3pt]north west:\NP}] (NP) at (0,0) {};
	\node[left = of NP,xshift=-.7cm,yshift=0.1cm] (SAT) {\color{black}  Circuit non-equivalence};
		\draw[-Circle,shorten >= -.18cm] (SAT) -- (-0.6,0.2) (NP);
	\node [e = blue!80!white, opacity=0.5, minimum height=1cm] (BQP) at (.8,-0.1) {~~~~~~~~~~~~~~~~~~~};
	\node at (1.7,-0.05) {\color{white}\BQP};

  \node[below right = of BQP, yshift=.6cm, xshift=-.2, text width=5.9cm] (QCS) {\color{black} Quantum circuit simulation (sampling)};
  \draw[-Circle,shorten >= -.18cm] (QCS) -- (BQP);

	\node [e = blue!60!white, opacity=.7,text opacity=1, minimum height=.98cm] at (.4,-0.1) {\color{white}~~~~~~\BPP};

  \node [venn circle = blue!80,text width=.3cm,align=center, minimum width=.3cm,opacity=.5,text opacity=1, xshift=-0.2cm] (P) at (.2,-0.1) {{\color{white}~\P}};  
  \node[below left = of P,yshift=.5cm] (CVP) {\color{black} Circuit evaluation (simulation)};
  \draw[-Circle,shorten >= -.18cm] (CVP) -- (P);
\end{tikzpicture}					
}
}

We finish by giving some references for further reading: Lipton and Regan's accessible introduction to quantum computing for computer scientists~\cite{lipton2021introduction}, a seminal textbook by Nielsen and Chuang~\cite{nielsen2000quantum}, the lecture notes of Ronald de Wolf~\cite{de2019quantum} and Watrous' detailed monograph on quantum information~\cite{watrous2018theory}.

\subsubsection*{Acknowledgements. } This publication is part of the project Divide \& Quantum  (with project number 1389.20.241) of the research program NWA-ORC which is (partly) financed by the Dutch Research Council (NWO). This work was supported
by the Dutch National Growth Fund, as part of the Quantum Delta NL program.
The third author acknowledges the support received through the NWO Quantum Technology program (project number NGF.1582.22.035).

\bibliographystyle{splncs04}
\bibliography{lit}

\newpage
\appendix
\section{Other Universal Gate Sets}
\label{app:gatesets}

In this section, we introduce the Clifford + $T$ gate set, which, like the Toffoli + $H$ gate set, is also universal for quantum computing. However, while the latter only requires real numbered amplitudes, the former also uses complex numbers, as is common in quantum mechanics.\footnote{While this is not relevant for the universality of quantum gate sets, some physical experiments indeed seem to require complex numbers~\cite{renou2021quantum}, an artifact of quantum mechanics that puzzled many scientists from the start.} 

The Clifford + $T$ gate set can be better understood when translating quantum states to a different basis, namely the so-called Pauli basis. We therefore discuss how to translate between the default computational basis and the Pauli, which requires us to first reinterpret states in the more general \concept{density matrix} formulation.

\subsection{The Clifford + $T$ Gate Set}
\label{appendix:Pauli+Clifford+T}

In \autoref{sect:quantum-info--gates}, some basic quantum gates were introduced: the $H, NOT, CNOT$ and $CCNOT$ ($\text{Toffoli}$) gates. In practice, often other quantum gates are used as well. Common families of quantum gates are the Pauli gates and the Clifford gates, which we will introduce now. 

The Pauli gates are given by the following matrices: 
\[
   I\, \equiv
    \begin{bmatrix}
      1 & 0 \\
      0 & 1
    \end{bmatrix},
    \quad
    X \equiv
    \begin{bmatrix*}[r]
      0 & 1 \\
      1 & 0
    \end{bmatrix*},
    \quad
     Y \equiv
    \begin{bmatrix*}[r]
      0 & -\dot{\imath} \\
      \dot{\imath} & 0
    \end{bmatrix*},
    \quad
    Z \equiv
    \begin{bmatrix*}[r]
      1 & 0 \\
      0 & -1
    \end{bmatrix*}
.
\]
One should note that the $X$ gate is equal to the $NOT$ gate. They are directly related to quantum physics (Wolfgang Pauli was a physicist). They are also used for quantum analysis. See, for example, \autoref{sec:densityMatricesInPauliBasis}.

An important subset of quantum gates is the \emph{Clifford gates},
which are widely used in error-correcting codes and can describe interesting quantum mechanical phenomena such as entanglement and teleportation.
A group of Clifford gates, or \emph{Clifford group}, is generated by the $H, CNOT$ and $S$ gates. We have not seen the $S$ gate before. It is defined as follows. \[S=\left[\begin{matrix}
    1 & 0 \\
    0 & i \\   
\end{matrix}\right].\] Circuits consisting only of Clifford gates can be classically simulated efficiently according to the Gottesman-Knill theorem~\cite{gottesman1998heisenberg}.

To create a universal gate set, we merely need to add one non-Clifford gate to the Clifford gates. Typically, the $T$ gate is used for this purpose, it is defined as follows.
 \[T = \left[\begin{matrix}
    1 & 0 \\
    0 & e^{i\pi/4} \\   
\end{matrix}\right],\]
The resulting Clifford + $T$ gate set is universal for quantum computing as shown by the Solovay--Kitaev theorem~\cite{dawson2005solovaykitaev}. This theorem shows that any quantum circuit on $n$ qubits can be approximated by a $\poly(n)$-sized circuit consisting of Clifford + $T$ gates. Therefore, the Clifford + $T$ gate set is important in quantum computing.

\subsection{Writing quantum states as density matrices in the Pauli basis}
\label{sec:densityMatricesInPauliBasis}
As is shown in \autoref{sec:quantumComputingFromCSperspective},
in an $n$-qubit quantum system,
a quantum state $\ket{\psi}$ is represented as a $2^n$ dimensional column vector.
We can also represent a quantum state $\ket{\psi}$ as its $2^n \times 2^n$ density matrix $\ket{\psi} \cdot \bra{\psi}$,
where $\bra\psi$ represents the conjugate transpose $\ket\psi^\dagger$ of $\ket\psi$~\cite{nielsen2000quantum}.
A quantum gate $G$ transforms the density matrix $\op{\psi}{\psi}$ to $G\op{\psi}{\psi}G^\dagger$, an action called \emph{conjugation}.
\begin{example}\label{ex:conjugation}
Given a state $\ket{0}$,
after performing gate $H$,
we get $\ket + \defn H\ket{0} = \frac{1}{\sqrt{2}}(\ket{0} + \ket{1})$.
In the density matrix picture, this computation is as follows:
\[
H (\op{0})H^\dagger = 
\left(\frac{1}{\sqrt{2}}\begin{bmatrix}
1 & 1 \\
1 & -1  
\end{bmatrix}\right)
\begin{bmatrix}
1 & 0 \\
0 & 0  
\end{bmatrix}
\left(\frac{1}{\sqrt{2}}
\begin{bmatrix}
1 & 1 \\
1 & -1  
\end{bmatrix}\right)
=
\frac{1}{2}
\begin{bmatrix}
1 & 1 \\
1 & 1  
\end{bmatrix}
=
\op+
\]

\vspace{-2.3em}
~\hfill$\diamond$
\end{example}
The Pauli matrices form a basis for the space of $2 \times 2$ matrices, which means that any such matrix, and in particular any single-qubit density matrix $\rho$, can be decomposed as $\rho = \lambda_0 I + \lambda_1 X + \lambda_2 Y + \lambda_3 Z$; we will call such a decomposition the \emph{Pauli decomposition} of $\rho$.
It is well known that, in general, the $\lambda_j$ are complex numbers, but if $\rho$ represents a density matrix, they are real (proof in \cite{mei2024eq}).
When generalizing to $n$ qubits,
a density matrix $\rho$ can be decomposed as $\rho = \sum_{P\in\texttt{PAULI}_n}\lambda_P \cdot P$,
where we call $\lambda_P$ the \emph{Pauli coefficient} of $P$, which is \emph{Pauli string}: a Kronecker product of $n$ Pauli matrices, for example $X\otimes Z \otimes I$.
We write $\texttt{PAULI}_n$ for the set of Pauli strings.
Therefore by moving from the vector representation in computational basis to the density-matrix representation in the Pauli basis, the amplitudes become real-valued rather than complex numbers.

\begin{example}
Given a quantum state $\ket{i} = \frac{1}{\sqrt{2}}(\ket{0} + i\ket{1})$,
its density matrix is 
$\op{i}{i} = \frac{1}{2}
\begin{bmatrix}
1 & -i \\
i & 1  
\end{bmatrix}$,
which can be decomposed in Pauli basis as $\op{i}{i} = \frac{1}{2}(Y+I)$.
Note the complex numbers in the density matrix, while the Pauli coefficients are real.
\hfill$\diamond$
\end{example}

\begin{example}
We apply $H$ to $\ket{0}$, computed in the vector picture in \autoref{ex:conjugation}, in the density-matrix picture.
First, we decompose $\op{0}{0}$ in the Pauli basis:
\[
\op{0}{0} = 
\begin{bmatrix}
1 & 0 \\
0 & 0  
\end{bmatrix}
=
\frac{1}{2}
\begin{bmatrix}
1 & 0 \\
0 & -1  
\end{bmatrix}
+ 
\frac{1}{2}
\begin{bmatrix}
1 & 0 \\
0 & 1  
\end{bmatrix}
= \frac{Z+I}{2}.
\]
It is straightforward to compute the $H \cdot Z \cdot H^\dagger = X$ by matrix multiplication. 
Using this, we compute the density matrix of the state $H\ket{0}$ in the Pauli basis,
\[
H\op{0}{0}H^\dagger = H\frac{Z+I}{2}H^\dagger = \frac{HZH^\dagger + HIH^\dagger}{2} = \frac{X+I}{2} 
=
\frac{1}{2}\begin{bmatrix}
1 & 1 \\
1 & 1  
\end{bmatrix},
\]
which is indeed the density matrix of $\ket{+}$ from \autoref{ex:conjugation}.
\hfill$\diamond$
\end{example}
\begin{figure}[t!]
  \centering
    \begin{minipage}{.45\textwidth}
    \begin{center}
    \begin{tikzpicture}[
    scale=0.3,
    every path/.style={>=latex},
    every node/.style={},
    inner sep=0pt,
    minimum size=14pt,
    line width=1pt,
    node distance=.5cm,
    thick,
    font=\footnotesize
    ]

    \node[draw, circle, minimum size=0.9cm] (00)   {$\ket{0}$};
    \node[draw,circle, right =1cm of 00, xshift=0cm, minimum size=0.9cm] (10) {$\ket{1}$};

\draw[\bigarrowhead, bend left=20] (00) edge node[yshift=0.2cm] {$\nicefrac{1}{\sqrt{2}}$} (10);
\draw[loop left, \bigarrowhead] (00) edge node[yshift=0.2cm, out=45, in=90] {$\nicefrac{1}{\sqrt{2}}$} (00);

\draw[\bigarrowhead, bend left=20] (10) edge node[yshift=-0.2cm] {$\nicefrac{1}{\sqrt{2}}$} (00);
\draw[loop right, \bigarrowhead] (10) edge node[yshift=0.2cm, out=45, in=90] {$-\nicefrac{1}{\sqrt{2}}$} (10);
\end{tikzpicture}
    \end{center}
    \end{minipage}
    \hspace{1cm}
    \begin{minipage}{.45\textwidth}
    \begin{center}
    \scalebox{1}{
\begin{tikzpicture}[
    scale=0.3,
    every path/.style={>=latex},
    every node/.style={},
    inner sep=0pt,
    minimum size=14pt,
    line width=1pt,
    node distance=.5cm,
    thick,
    font=\footnotesize
    ]

    \node[draw, circle, minimum size=0.9cm] (X)   {${X}$};
    \node[draw,circle, right =1cm of X, xshift=0cm, minimum size=0.9cm] (Y) {${Y}$};
    \node[draw,circle, below =1cm of X, xshift=0cm, minimum size=0.9cm] (Z) {${Z}$};
    \node[draw,circle, below =1cm of Y, xshift=0cm, minimum size=0.9cm] (I) {${I}$};

\draw[loop right, \bigarrowhead] (Y) edge node[yshift=0.2cm, out=45, in=90] {$-1$} (Y);
\draw[bend left=20, \bigarrowhead] (Z) edge node[xshift=-0.2cm, out=45, in=90] {$1$} (X);

\draw[loop right, \bigarrowhead] (I) edge node[yshift=0.2cm, out=45, in=90] {$1$} (I);
\draw[bend left=20, \bigarrowhead] (X) edge node[xshift=0.2cm, out=45, in=90] {$1$} (Z);
\end{tikzpicture}
}
    \end{center}
    \end{minipage}
  \caption{The effect on applying an $H$ gate on a single qubit, visualized as an automaton with computation-basis vectors (left) and the Pauli matrices (right) as nodes. For interpretation, see the text.}
  \label{fig:com2pauli}
\end{figure}
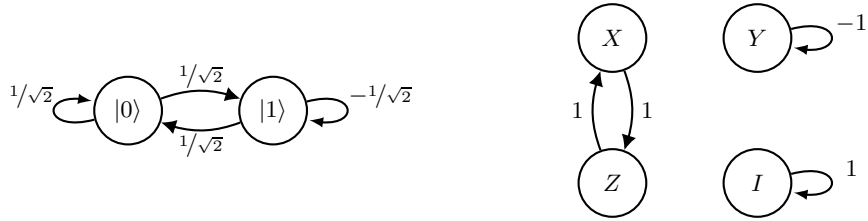

Extending the above example by conjugating $H$ gates on all Pauli matrices,
one can compute by matrix multiplication that
\[
HZH^\dag = X, \quad  HXH^\dag = Z, \quad HYH^\dag = -Y, \quad HIH^\dag = I.
\]
\autoref{fig:com2pauli} visualizes the action of the $H$ gate on the Pauli matrices, compared to computational-basis vectors (as done in \autoref{sec:prelims}).
We observe that the outdegree in the latter is 2, but it is only 1 in the former.
Consequently, in the Pauli basis, there is no increase in the number of paths from one automaton node to another, thereby avoiding the blow-up in the number of paths between automaton nodes that will be present for the computational basis if multiple Clifford gates are applied one after another.

The automaton states for the other two Clifford gates, $S$ and $CNOT$, also have out-degree 1.
We leave it as an exercise for the reader to draw these automata, which follow from the conjugation action of $S$ and $CNOT$ on Pauli strings, which are given in \autoref{tab:clifford}.

\keymessage[s]{
\begin{itemize}
    \item Quantum states can be represented as density matrices and these can be written as a linear combination of Pauli strings with real-valued weights.
    \item Suppose that a quantum state is represented by a density matrix, written as linear combination of Pauli matrices. Applying a gate to the quantum state yields a density matrix with the same number of Pauli matrices with nonzero Pauli coefficients
 (this follows from the outdegree of $1$ in the automaton).
\end{itemize}
}

\section{Simulating Quantum Circuits in the Pauli Basis}
\label{app:pauli}

In this section, we show how to reduce simulation of quantum computation to weighted model counting (weighted-\#SAT) in the Pauli basis (see \aref{app:gatesets}).
Examples and figures in this section are drawn from~\cite{mei2024simulating,mei2024eq}.

\subsection{A first step towards encoding quantum circuits using Boolean variables}

The Pauli basis is key to efficient simulation of Clifford circuits~\cite{gottesman1997stabilizer}, which relies on two facts:
\begin{enumerate}
\item first, a succinct representation of the Pauli decomposition of the input state;
\item second, the fact that the number of nonzero Pauli coefficients in the density matrix is invariant under Clifford gates (and that the resulting density matrix can still be represented succinctly).
\end{enumerate}
We will now show how one could reduce Clifford simulation to SAT by first providing a single Boolean formula encoding the input quantum state (item 1, in \autoref{sec:encoding-state}), followed by extending the formula to encode the quantum state's Clifford gate update (item 2, \autoref{sec:encoding-gate}).
We remark that this is silly on its own, since Clifford-circuit simulation is in the complexity class P while solving SAT is NP-complete.
However, this step will be necessary to move to reducing universal quantum computing to weighted-\#SAT later.

\subsubsection{Encoding the input quantum state}
For the first step, we use a common encoding of a Pauli matrix $\sigma[x,z]$ by two Boolean variables $x,z$ as
\[
  \begin{aligned}
   \sigma[00] \equiv \,I, \quad
    \sigma[01] \equiv Z, \quad
 \sigma[10] \equiv X, \quad
   \sigma[11]  \equiv Y .
  \end{aligned}
\]
As can be seen in \autoref{tab:clifford}, the Clifford gates map Pauli strings under conjugation to a Pauli string with factor $\pm 1$, hence we will also need to encode this factor.
We do so by adding a variable $r$ for the sign $(-1)^r$ of a Pauli matrix.
Let Boolean variables $x$ and $z$ encode Pauli matrices.
For example, the Boolean formula $r\neg xz$ encodes $-Z$ as its one satisfying assignment is $\{r\rightarrow 1, x\rightarrow 0, z\rightarrow 1\}\equiv - Z$.
The encoding can be naturally generalized to $n$-qubits Pauli strings,
which can be represented by $2n+1$ Boolean variables,
i.e., $x_j$ and $z_j$ for $j$-th qubit and $r$ for sign of the Pauli string.
We use this encoding to succinctly represent the input state $\ket{0}^{\otimes n}$ to a quantum circuit, as shown in the following example for $n=2$.

\begin{example}\label{ex:init}
  The density operator $\op{00}{00} = \frac{1}{4}(I\otimes I + I\otimes Z + Z\otimes I + Z\otimes Z)$ is represented by the formula $\neg r\neg x_{0}\neg x_{1}$, as its satisfying assignments are:
  \[
  \left\{
    \begin{aligned} 
    &\{r \rightarrow 0, x_0 \rightarrow 0, x_1\rightarrow 0, z_0\rightarrow 1, z_1\rightarrow 1\} \\
    &\{r \rightarrow 0, x_0 \rightarrow 0, x_1\rightarrow 0, z_0\rightarrow 1, z_1\rightarrow 0\} \\
    &\{r \rightarrow 0, x_0 \rightarrow 0, x_1\rightarrow 0, z_0\rightarrow 0, z_1\rightarrow 1\} \\
    &\{r \rightarrow 0, x_0 \rightarrow 0, x_1\rightarrow 0, z_0\rightarrow 0, z_1\rightarrow 0\}
    \end{aligned}
     \right\}  \equiv 
  \left\{
    \begin{aligned} 
    Z \otimes Z\\
    Z \otimes I\\
    I \otimes Z\\
    I \otimes I
    \end{aligned}
     \right\} 
.
  \]
The factor $\frac{1}{4}$ of the density matrix need not be included in the encoding as one can derive that the $n$-qubit density matrix $\ket{0}^{\otimes n} \bra{0}^{\otimes n}$ has factor $\frac{1}{2^n}$.
\hfill$\diamond$
\end{example}

\subsubsection{Encoding the gate update}

\begin{table}[t!]
  \centering
  \caption{The action of conjugating a Pauli string by Clifford gates $H, S, CNOT$. For the controlled-NOT gate, subscripts indicate `control' and `target' qubit. Adapted from~\cite{garcia2014geometry}.}
  \label{tab:clifford}
  \setlength{\tabcolsep}{5pt} 
  \begin{tabularx}{0.65\textwidth}{c|rr||c|cc}
      \toprule
      \textbf{Gate} & \textbf{In} & \textbf{Out} & \textbf{Gate} & \textbf{In} & \textbf{Out} \\
      \midrule
      & $X$ & $Z$ & \multirow{6}{*}{CNOT} & $\phantom{-}I_c \otimes X_t$ & $\phantom{-}I_c\otimes X_t$ \\
      $H$ & $Y$ & $-Y$ & & $\phantom{-}X_c \otimes I_t$ & $\phantom{-}X_c \otimes X_t$ \\
      & $Z$ & $X$ & & $\phantom{-}I_c \otimes Y_t$ & $\phantom{-}Z_c \otimes Y_t$ \\
      \cline{1-3}
       & $X$ & $Y$ & & $\phantom{-}Y_c \otimes I_t$ & $\phantom{-}Y_c \otimes X_t$ \\
      $S$ & $Y$ & $-X$ & & $\phantom{-}I_c \otimes Z_t$ & $\phantom{-}Z_c \otimes Z_t$ \\
      & $Z$ & $Z$ & & $\phantom{-}Z_c \otimes I_t$ & $\phantom{-}Z_c \otimes I_t$ \\
      \bottomrule
  \end{tabularx}
\end{table}

We have seen before that all $n$-qubit Pauli strings, with potential factor $-1$, can be encoded by $2n + 1$ Boolean variables.
We have also seen that a Clifford gate maps a Pauli string to $\pm 1$ times a Pauli string under conjugation. 
Combining these two facts brings us to a formula which describes how a Pauli string, represented by variables $(r^t, x^t, z^t)$ is mapped to a Pauli string $(r^{t+1}, x^{t+1}, z^{t+1})$ (where superscript $t$ indicates the time-step in the circuit).
For example, for the $H$ gate, its action on Pauli strings as given in \autoref{tab:clifford}, is translated to the Boolean-variable representation of Pauli strings in \autoref{tab:boolhgate}, which in turn is translated to the following formula:
\[
  F_{H^t} \defn r^{t+1} \Longleftrightarrow r^{t} \oplus x^t z^t
  ~~\land~~  z^{t+1} \Longleftrightarrow  x^t
  ~~\land~~  x^{t+1} \Longleftrightarrow  z^t.
\]
Similarly, we obtain Boolean formulas describing the update of a Pauli string under the Clifford gates $S$ and $CNOT$ based on the \autoref{tab:clifford}.
\begin{example}
Suppose that a $CNOT$ gate is the $41$st gate in a two-qubit circuit, where the first qubit (index $0$) is the control qubit and the second qubit (index $1$) the target.
The corresponding formula is
 \begin{equation*}
   \begin{aligned}
     F_{CNOT^{42}_{0,1}} \defn  ~& r^{43} \Longleftrightarrow r^{42} \oplus x^{42}_{0}z^{42}_{1}
       (x^{42}_{1} \oplus \neg z^{42}_{0})
     \
     ~~\land~~  x^{43}_{1} \Longleftrightarrow  x^{42}_{1} \oplus  x^{42}_{0} ~~\land~~ \\
     &  z^{43}_{0} \Longleftrightarrow  z^{42}_{0} \oplus  z^{42}_{1}  ~~\land~~ x^{43}_{0} \Longleftrightarrow  x^{42}_{0}
     ~~\land~~ z^{43}_{1} \Longleftrightarrow  z^{42}_{1}.
   \end{aligned}
 \end{equation*}
 
\vspace{-2.8em}
\hfill$\diamond$
\end{example}

\begin{table}[t!]
\caption{Boolean variables under the action of conjugating one $H$ gate.}
\label{tab:boolhgate}
\setlength{\tabcolsep}{6pt} %
\renewcommand{\arraystretch}{1} %
  \centering
  \begin{tabular}{rc|c|c|c|c}
    \hline
     $P$  & $x^t z^t r^t$ & $HPH^\dagger$  & $r^{t+1}$   & $x^{t+1}$ & $z^{t+1}$ \\
     \hline
     $I$    & 00 $r^t$ & $I$  & \multirow[]{3}{*}{$r^t$} & \multirow[]{4}{*}{$z^t$} & \multirow[]{4}{*}{$x^t$}  \\
    \cline{1-3}
    $Z$     & 01 $r^t$ &$X$ & $r^t$ &    &  \\
    \cline{1-3}
    $X$    & 10 $r^t$ & $Z$ &  &  &  \\
    \cline{1-4}
    $Y$     & 11 $r^t$ &$-Y$ & $\neg r^t$ &  &  \\
    \hline
  \end{tabular}
\end{table}

\keymessage{
Given a circuit with only Clifford gates, we can construct a Boolean formula whose single satisfying assignments represent the circuit's output quantum state.
Specifically, each assignment encodes a single Pauli matrix in the density matrix's Pauli decomposition which has a nonzero Pauli coefficient.
}

\subsection{How a $T$ gate acts on Pauli strings}
Any quantum circuit can be approximated by a circuit which has only Clifford gates and $T$ gates.
Having treated the Clifford gates above, we only need to encode the conjugation action of $T$ in Boolean variables.
However, doing so is not a straightforward extension of the Clifford case.
To see this, we compute that
\begin{equation}\label{eq:branch}
  TXT^\dag = \frac{1}{\sqrt{2}}(X+Y) \quad \text{and} \quad TYT^\dag=\frac{1}{\sqrt{2}}(Y-X).
\end{equation}

 \begin{wrapfigure}[9]{r}{7cm}
  \centering
\vspace{-1cm}
  \scalebox{0.85}{ 
  \begin{tikzpicture}[
    scale=0.3,
    every path/.style={>=latex},
    every node/.style={},
    inner sep=0pt,
    minimum size=14pt,
    line width=1pt,
    node distance=.5cm,
    thick,
    font=\footnotesize
    ]

    \node[draw, circle, minimum size=0.9cm] (X)   {${X}$};
    \node[draw,circle, right =1cm of X, xshift=0cm, minimum size=0.9cm] (Y) {${Y}$};
    \node[draw,circle, below =.5cm of X, xshift=0cm, minimum size=0.9cm] (Z) {${Z}$};
    \node[draw,circle, below =.5cm of Y, xshift=0cm, minimum size=0.9cm] (I) {${I}$};

\draw[loop left, \bigarrowhead] (X) edge node[out=45, in=90] {$\frac{1}{\sqrt{2}}$} (X);
\draw[bend left=20,\bigarrowhead] (X) edge node[yshift=0.3cm,  out=45, in=90] {$\frac{1}{\sqrt{2}}$} (Y);
\draw[bend left=20,\bigarrowhead] (Y) edge node[yshift=-0.3cm,  out=45, in=90] {$\frac{1}{\sqrt{2}}$} (X);

\draw[loop right, \bigarrowhead] (Y) edge node[yshift=0.2cm, out=45, in=90] {$\frac{1}{\sqrt{2}}$} (Y);
\draw[loop left, \bigarrowhead] (Z) edge node[out=45, in=90] {$1$} (Z);

\draw[loop right, \bigarrowhead] (I) edge node[yshift=0.2cm, out=45, in=90] {$1$} (I);
\end{tikzpicture}}
  \caption{Automaton which visualizes the conjugation action of the $T$ gate Pauli basis.}
  \label{fig:com2t}
\end{wrapfigure}
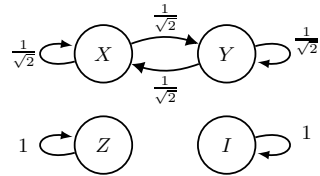
The transition system in \autoref{fig:com2t} visualizes the fact that a $T$ gate maps a single Pauli string to a linear combination of two Pauli strings for $X$ and $Y$: the automaton nodes have out-degree 2 in those cases, in contrast to the Clifford gates in \autoref{fig:com2pauli}.
The consequence of this is that the number of nonzero Pauli coefficients that need to be tracked when simulating a circuit of Clifford and $T$ gates might become exponentially large with the number of $T$ gates.

Furthermore,
when doing two $T$ gates in a row, i.e.
\[
  TTXT^\dag T^\dag = T(\tfrac{1}{\sqrt{2}}(X+Y))T^\dag = 
 \cancel{\tfrac{1}{2}(X-X)} + \tfrac{1}{2}(Y+Y) = Y,
\]
both constructive interference $(Y+Y)$ and destructive interference $(X-X)$ happen.
To capture those real-valued weights, we need more than merely Boolean variables, and we resort to weighted \#SAT.

\keymessage{
    The $T$ gate maps, under conjugation, a Pauli string to a summation of Pauli strings with non-trivial real-valued weights.
}

\subsection{An encoding of quantum circuits with weighted \#SAT}
Since quantum circuit compilation comprises many computationally hard reasoning tasks that 
nonetheless lie inside \#$\P$,
it is natural to consider to encode quantum computing in \#SAT,
which counts all the satisfying assignments.
We will now do so.

Quantum states in Pauli basis have real numbers for weights of Pauli strings rather than Booleans,
thus we consider introducing weights for the variables.
To deal with the interference, 
we restore the nontrivial weights of the Pauli string into the weights of newly introduced variables $u$.
We set $u=1$ to indicate there is a branch in the Pauli matrices as shown in \autoref{eq:branch}.
\autoref{tab:booltgate} illustrates the details of how the Boolean variables change over a $T$ gate. 
Given a $T$ gate on a single-qubit circuit at time step $t$,
the Boolean formula of $T^t$ can be obtained  as:
\begin{equation}\label{cons:cliffordt}
  \begin{aligned}
    F_{T^t} \defn  ~&x^{t+1} \Longleftrightarrow x^t \quad \wedge \quad 
  x^{t} \lor (z^{t+1} \Leftrightarrow z^t) \quad \wedge\\
  & r^{t+1} \Longleftrightarrow r^t \oplus  x^t  z^t  \neg z^{t+1}  \quad \wedge \quad u_{t} \Longleftrightarrow x^t.
  \end{aligned}
\end{equation}
where we define the weight function $W$ as $W(r^m) = -1$, $W(\no r^m) = 1$, $W(u_{t}) = \frac{1}{\sqrt{2}}$, $W(\no u_{t}) = 1$ for all $t\in\{0,\dots,m\}$
(and all other variables are unbiased).

\begin{example}
  The $T$ gate maps the state $\tfrac{1}{2}(I+Y)$ to $\tfrac{1}{2}T(I+Y)T^\dag  = \frac{1}{2}(I+\tfrac{1}{\sqrt{2}}(Y-X))$ under conjugation. %
  This is a linear combination of three Pauli strings and is represented by the following three satisfying assignments
  \[
    \begin{aligned}
      & \{r \rightarrow 0, x\rightarrow 0, z\rightarrow 0, u\rightarrow 0\} \equiv I \\
      & \{r \rightarrow 1, x\rightarrow 1, z\rightarrow 0, u\rightarrow 1\} \equiv - \tfrac{1}{\sqrt{2}}X \\
      & \{r \rightarrow 0, x\rightarrow 1, z\rightarrow 1, u\rightarrow 1\} \equiv \tfrac{1}{\sqrt{2}}Y
    \end{aligned}
  \]
  
\vspace{-2.2em}
\hfill$\diamond$
\end{example}

At this point, we have all ingredients to simulate a sequence of Clifford and $T$ gates by weighted \#SAT by Boolean constraint $F_C$ with weight function $W$,
where the output state is
\[
\rho = \sum_{\alpha\in SAT(F_C)} W(\alpha(r^m))\prod_{t\in[m]}W(\alpha(u_t))\sigma[\alpha(x), \alpha(z)].
\]

\begin{table}[t!]
  \caption{Boolean variables under the action of conjugating one T gate.}
  \label{tab:booltgate}
  \setlength{\tabcolsep}{6pt} %
  \renewcommand{\arraystretch}{1} %
    \centering
    \begin{tabular}{rc|c|c|c|c|c}
      \hline
       $P$  & $x^t z^t r^t$ & $TPT^\dag$  & $x^{t+1}$ & $z^{t+1}$ & $r^{t+1}$ & $u$  \\
       \hline
       $I$    & 00 $r^t$ & $I$ & \multirow[]{2}{*}{0} & \multirow[]{2}{*}{$z^t$} & \multirow[]{2}{*}{$r^t$} &\multirow[]{2}{*}{0} \\
  \cline{1-3}
      $Z$     & 01 $r^t$ &$Z$  &  & &  \\
      \hline
      $X$    & 10 $r^t$ &$\frac{1}{\sqrt{2}}(X+Y)$ & \multirow[]{2}{*}{1} & \multirow[]{2}{*}{\{0,1\}} & $r^t$ & \multirow[]{2}{*}{$1$} \\
      \cline{1-3}\cline{6-6}
      $Y$     & 11 $r^t$ &$\frac{1}{\sqrt{2}}(Y-X)$ & & & $r^t\oplus \neg z^{t+1}$ & \\
      \hline
    \end{tabular}
  \end{table}
\begin{example}\label{ex:clifftwmc}
Consider the following circuit on two qubits: 
    \[
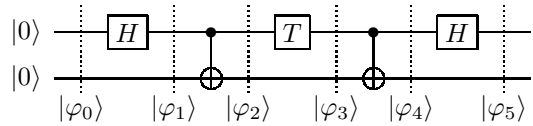
     \begin{array}{c}  
       \Qcircuit @C=1em @R=.7em {
         \lstick{\ket{0}} & \qw\ar@{.}[]+<0em,1em>;[d]+<0em,-0.5em> & \gate{H} &\qw\ar@{.}[]+<0em,1em>;[d]+<0em,-0.5em> & \ctrl{1} & \qw\ar@{.}[]+<0em,1em>;[d]+<0em,-0.5em> & \gate{T} &\qw\ar@{.}[]+<0em,1em>;[d]+<0em,-0.5em> & \ctrl{1} & \qw\ar@{.}[]+<0em,1em>;[d]+<0em,-0.5em> &\gate{H} &\qw\ar@{.}[]+<0em,1em>;[d]+<0em,-0.5em> &\qw \\
         \lstick{\ket{0}} & \qw & \qw & \qw & \targ & \qw &\qw&\qw &\targ & \qw &\qw& \qw &\qw\\
         & \ket{\varphi_0} & & \ket{\varphi_1} & & \ket{\varphi_2} & & \ket{\varphi_3} & & \ket{\varphi_4} & & \ket{\varphi_5} &
       }
   \end{array}
   \]
Let $F_{\init (\ket{00})} = \neg r \neg x_0 \neg x_1$ encoding $\ket{00}$ as given in \autoref{ex:init}.
We will now derive the Pauli decomposition of the output state $\ket{\varphi_5}$ in two ways.
First, using weighted model counting: the formula is $F_{\init (\ket{00})}\wedge F_{H_0^0}\wedge F_{CNOT_{0,1}^1} \wedge F_{T_0^2} \wedge F_{CNOT_{0,1}^3} \wedge F_{H_0^4}$\footnote{The subscript(s) of the gate indicate the qubit(s) the gate acts on.} %
with satisfying assignments:
\begin{equation}
\label{eq:ex-clifftwmc}
    \left\{
        \begin{aligned}
            & \{r^5 \rightarrow 0, x^5_0 \rightarrow 0, x^5_1\rightarrow 0, z^5_0\rightarrow 1, z^5_1\rightarrow 0, u_2\rightarrow 1\}, \\
            & \{r^5\rightarrow 1, x^5_0 \rightarrow 1, x^5_1\rightarrow 0, z^5_0\rightarrow 1, z^5_1\rightarrow 0, u_2\rightarrow 1\}, \\
            & \{r^5 \rightarrow 0, x^5_0 \rightarrow 0, x^5_1\rightarrow 0, z^5_0\rightarrow 0, z^5_1\rightarrow 1, u_2\rightarrow 0\}, \\
            & \{r^5 \rightarrow 0, x^5_0 \rightarrow 0, x^5_1\rightarrow 0, z^5_0\rightarrow 1, z^5_1\rightarrow 1, u_2\rightarrow 1\}, \\
            & \{r^5 \rightarrow 1, x^5_0 \rightarrow 1, x^5_1\rightarrow 0, z^5_0\rightarrow 1, z^5_1\rightarrow 1, u_2\rightarrow 1\}, \\
            & \{r^5 \rightarrow 0, x^5_0\rightarrow 0, x^5_1\rightarrow 0, z^5_0\rightarrow 0, z^5_1\rightarrow 0, u_2\rightarrow 0\}
        \end{aligned}
      \right\}
      \equiv 
  \left\{
    \begin{aligned} 
    \tfrac{1}{\sqrt{2}} Z \otimes I\\
   -\tfrac{1}{\sqrt{2}} Y \otimes I\\
    I \otimes Z\\
    \tfrac{1}{\sqrt{2}} Z \otimes Z\\
   -\tfrac{1}{\sqrt{2}} Y \otimes Z\\
    I \otimes I
    \end{aligned}
     \right\} 
\end{equation}
  where $W(r^5) = 1$, $W(\no r^5) = -1$, $W(u_2) = \frac{\sqrt{2}}{2}$ and $W(\no u_2) = 1$.
  Here we omit the satisfying assignments for $\{r^t,x^t_0,x^t_1,z^t_0,z^t_1 \mid 0\leq t\leq 4\}$.
We now translate the satisfying constraints (left-hand side of \autoref{eq:ex-clifftwmc}) into Pauli matrices with Pauli coefficients (right-hand side of \autoref{eq:ex-clifftwmc}.
From the latter, we compute the density matrix:
\begin{eqnarray}
\nonumber
\op{\varphi_5}{\varphi_5} &= \frac{1}{4}(\tfrac{1}{\sqrt{2}} Z \otimes I -\tfrac{1}{\sqrt{2}} Y \otimes I + I \otimes Z + \tfrac{1}{\sqrt{2}} Z \otimes Z -\tfrac{1}{\sqrt{2}} Y \otimes Z + I \otimes I) \\
\nonumber
&= \frac{1}{4}\left(
\tfrac{1}{\sqrt{2}}
\left[\begin{smallmatrix}
    1 & 0 & 0 & 0 \\
    0 & 1 & 0 & 0 \\
    0 & 0 & -1 & 0 \\
    0 & 0 & 0 & -1 \\
\end{smallmatrix}\right]
- \tfrac{1}{\sqrt{2}}
\left[\begin{smallmatrix}
    0 & 0 & -i & 0 \\
    0 & 0 & 0 & -i \\
    i & 0 & 0 & 0 \\
    0 & i & 0 & 0 \\
\end{smallmatrix}\right]
+ 
\left[\begin{smallmatrix}
    1 & 0 & 0 & 0 \\
    0 & -1 & 0 & 0 \\
    0 & 0 & 1 & 0 \\
    0 & 0 & 0 & -1 \\
\end{smallmatrix}\right]\right. \\
& \quad + \tfrac{1}{\sqrt{2}}
\left[\begin{smallmatrix}
    1 & 0 & 0 & 0 \\
    0 & -1 & 0 & 0 \\
    0 & 0 & -1 & 0 \\
    0 & 0 & 0 & 1 \\
\end{smallmatrix}\right]
- \tfrac{1}{\sqrt{2}}
\left[\begin{smallmatrix}
    0 & 0 & -i & 0 \\
    0 & 0 & 0 & i \\
    i & 0 & 0 & 0 \\
    0 & -i & 0 & 0 \\
\end{smallmatrix}\right]
+ 
\left.\left[\begin{smallmatrix}
    1 & 0 & 0 & 0 \\
    0 & 1 & 0 & 0 \\
    0 & 0 & 1 & 0 \\
    0 & 0 & 0 & 1 \\
\end{smallmatrix}\right]
\right) = \left[\begin{smallmatrix}
    \frac{1}{2} + \frac{\sqrt{2}}{4} & 0 & \frac{\sqrt{2}}{4}i & 0 \\
    0 & 0 & 0  & 0 \\
    -\frac{\sqrt{2}}{4}i & 0 & \frac{1}{2} - \frac{\sqrt{2}}{4}  & 0 \\
    0 & 0 & 0 & 0
\end{smallmatrix}\right].
\label{eq:big-dm}
\end{eqnarray}

Alternatively, using the vector representation in the computational basis and following \autoref{sec:prelims}, one can derive (exercise to the reader) that the vector representing the circuit's output state must be
\begin{equation}
\label{eq:phifive}
\ket{\varphi_5} = 
\left[
\frac{1}{2}+ \frac{1}{2}\exp(i\frac{\pi}{4}),0, \frac{1}{2} - \frac{1}{2}\exp(i\frac{\pi}{4}), 0
\right]^{\top}
\end{equation}
whose density matrix is precisely the one in \autoref{eq:big-dm}.
\hfill $\diamond$
\end{example}

\keymessage{
Clifford circuits equipped with a non-Clifford gate, e.g. $T$ gate, enable universal quantum computing.
Simulating such circuits can be reduced to weighted model counting.
}

\subsection{Encoding measurements}
Above, we showed, given a quantum circuit, how to construct a Boolean formula whose weighted model count encodes the output quantum state of the circuit.
Now we explain how to encode a measurement of the most significant qubit at the end of the circuit.
To find this measurement's outcome probabilities, we use the result that the probability of outcome $0$ equals $p_0 = \frac{1}{2}(1 + \lambda_{Z_0})$ if the measured state is $\rho = \sum_{P\in\texttt{PAULI}_n} \lambda_P P$,  where $Z_0$ is shorthand for $Z \otimes I^{\otimes n - 1}$
(seeing why this is the case requires more information than given in this tutorial; we refer to \cite{mei2024simulating} for a proof).
But obtaining $\lambda_{Z_0}$ is easy using weighted model counting: given a Boolean formula $F_{\rho}$ for the output state, we conjunct it with a constraint whose only satisfying assignment is $Z_0$:
\begin{equation}\label{cons:measure}
  F_{M}(V) \defn \bigwedge_{j\in[n]} \neg x_j \wedge \bigwedge_{j\in[n], j\neq 0} \neg z_j \wedge z_0 ~ \equiv ~ Z_0.
\end{equation}
With \autoref{cons:measure}, the satisfying assignments for $Z_0$ will be left in the final result while all other assignments are ruled out.

\begin{example}\label{ex:mea}
  Reconsider in \autoref{ex:clifftwmc},
  we can obtain the probability of outcome $0$ when measuring the first qubit of $\ket{\varphi_5}$
  by adding the measurement constraint $F_M \defn \no x^5_0 \wedge \no x^5_1 \wedge z^5_0 \wedge \no z^5_1$ to the circuits constraints.
  The constraint $F_M$ collects all amplitudes of $Z\otimes I$ in the output state; thus, the resulting weight is to sum them up.
  The satisfying assignments will be the subset of the solutions in \autoref{ex:clifftwmc}:
$
  \left\{\sigma = \{r^5 \rightarrow 0, x^5_0 \rightarrow 0, x^5_1\rightarrow 0, z^5_0\rightarrow 1, z^5_1\rightarrow 0, u_2\rightarrow 1\}\right\}
$
  where we only show the variables representing the final state.
  The resulting probability is
  $\frac{1}{2} + \frac{1}{2}W(\sigma(r^5))W(\sigma(u_2)) = \frac{1}{2} + \frac{\sqrt{2}}{4}$.

This probability is indeed equal to the probability of outcome 0 that one computes by using the state vector 
\autoref{eq:phifive} directly: $|\frac{1}{2}+ \frac{1}{2}\exp(i\frac{\pi}{4})|^2 + |0|^2 = \frac{1}{2}+\frac{\sqrt{2}}{4}$.
\hfill$\diamond$
\end{example}

To summarize, the above \#SAT encoding relies on two crucial steps. First, transforming states in the computational basis to density matrices in the Pauli basis alleviates the need for complex numbers, making the method more suitable for existing, modern model counters. Second, by using the weighted satisfying assignments to represent individual components of the Pauli basis (i.e., Pauli strings with real weights), the encoding stays linear in the size of the circuits, as shown by the fact that each gate only requires a short constraint that can easily be expressed as a constant number of clauses. The exponential blowup is restrained to where it belongs: inside the heuristic search algorithms in the \#SAT solver, which can often efficiently deal with it.
The encoding has been shown to outperform existing techniques in some cases~\cite{mei2024simulating} and has also been applied to the different task of checking whether two quantum circuits with gates only represent the same unitary operator (circuit equivalence checking)~\cite{mei2024eq}.

\end{document}